\documentstyle[prl,aps,psfig]{revtex}
 
\def\be{\begin{equation}}
\def\ee{\end{equation}}
\def\bea{\begin{eqnarray}}
\def\eea{\end{eqnarray}}
\def\ben{\begin{eqnarray}}
\def\een{\end{eqnarray}}
\def\>{\rangle}
\def\<{\langle}

\def\ep{\epsilon}

\newcommand{\ket}[1]{{|#1\rangle}}
\newcommand{\bra}[1]{{\langle#1|}}

\newcommand{\trace}{\mbox{tr}}

\newcommand{\Exp}{\mbox{Exp}}
\newcommand{\defeq}{\doteq}

\newcommand{\mod}{\;\mbox{mod}}

\renewcommand{\tensor}{\otimes}

\def\cS{{\cal S}}
\def\cL{{\cal L}}

\def\cN{{\cal N}}
\def\cP{{\cal P}}
\def\cT{{\cal T}}
\def\cD{{\cal D}}
\def\cR{{\cal R}}
\def\cH{{\cal H}}

\begin{document}

\title{Effective Pure States for Bulk Quantum Computation}

\author{E. Knill$^1$, I. Chuang$^2$, R. Laflamme$^2$}

\address{\vspace*{1.2ex}
        \hspace*{0.5ex}{$^1$ Computer Research and Applications CIC-3,
 	MS B-265;} \hspace*{0.5ex}{$^2$ Theoretical Astrophysics T-6,
 	MS B-288 \\ Los Alamos National Laboratory, Los Alamos, NM
 	87455 }}

\date{\today}

\maketitle
\begin{abstract}
In bulk quantum computation one can manipulate a large number of
indistinguishable quantum computers by parallel unitary operations and
measure expectation values of certain observables with limited
sensitivity. The initial state of each computer in the ensemble is
known but not pure. Methods for obtaining effective pure input states
by a series of manipulations have been described by Gershenfeld and
Chuang~\cite{chuang:qc1996c,chuang:qc1997a} (logical labeling) and
Cory et al.~\cite{cory:qc1996a,cory:qc1997a} (spatial averaging) for
the case of quantum computation with nuclear magnetic resonance. We
give a different technique called temporal averaging. This method is
based on classical randomization, requires no ancilla qubits and can
be implemented in nuclear magnetic resonance without using gradient
fields. We introduce several temporal averaging algorithms suitable
for both high temperature and low temperature bulk quantum computing
and analyze the signal to noise behavior of each.
\end{abstract}



\section{Introduction}

Quantum computation involves the transformation of one known pure
quantum state into another unknown state, which can be measured to
provide a computationally useful output.  Traditionally, it has been
understood that an important part of this process is proper
preparation of a fiducial initial pure state, such that the
computational input is well known, and the output is thus meaningful.
In particular, it has usually been assumed that the input cannot be a
stochastic mixture.  However, two
groups\cite{chuang:qc1996c,chuang:qc1997a,cory:qc1996a,cory:qc1997a}
have recently shown that by using a different technique, called {\em
bulk quantum computation}, the same computation can be performed but
with an initial mixture state, which is often much easier to achieve
experimentally.  Bulk quantum computation is being implemented for
small numbers of qubits using nuclear magnetic resonance (NMR)
techniques.

Bulk quantum computation is performed on a large ensemble of indistinguishable
quantum computers. At the beginning of a computation, each member $c$ of the
ensemble is in an initial state $\rho_{c,0}$ such that the average $\rho_{0}
\defeq \Exp(\rho_{c,0})$ of these states is known. A bulk computation with
such an ensemble can be divided into three steps consisting of preparation,
computation and readout. Each of these steps is equivalent to an application
of the same quantum operation to each member of the ensemble. The purpose of
the preparation step is to transform the input state to an {\em effective pure
state} which permits an unbiased observation of the output of the algorithm.
The computation is assumed to be a fixed unitary operator derived from a
standard quantum algorithm, that is an algorithm with a one qubit answer.  We
wish to determine this answer on input $\ket{\mbox{\bf 0}}$ (the state where
every qubit is $\ket{0}$).  The readout procedure may include some
postprocessing of the algorithm's output and terminates in the measurement of
the observable $\sigma_z^{(1)}$, the spin along the $z$-axis of the first
qubit. In bulk quantum computation, the measurement yields a noisy version of
the average value of $\sigma_z^{(1)}$ over the ensemble of quantum
computers. For our signal to noise analyses, we assume that the noise is
unbiased with variance $s^2$.

Formally, a bulk quantum computation of an algorithm implementing the unitary
transformation $C$ with preparation and postprocessing operations $\cP$ and
$\cR$ transforms $\rho_{0}$ to
\bea
\rho_{out} &=& \sum_{i,j}R_iCP_j\rho_{0}P_j^\dagger C^\dagger R_i^\dagger,
\eea
where the $R_i$ and $P_j$ are the operators in a linear representation
of the quantum operations $\cP$ and $\cR$~\cite{schumacher:qc1996a}.
The measurement step of the readout procedure yields
$\trace(\rho_{out}\sigma_z^{(1)})$ with noise.  In the methods
investigated in this paper, $\cR$ is unitary, usually the
identity. The purpose of $\cP$ is to create an {\em effective pure
state\/}. The simplest example of an effective pure
state\footnote{Cory et al.~\cite{cory:qc1996a,cory:qc1997a} call this
a {\em pseudo-pure state\/}.} is a density matrix of the form
\ben
\sum_j P_j\rho_{0} P_j^\dagger &=&
  p\,\ket{0}\bra{0} + {q\over N} I.
\een
Here $N = \dim (I) = 2^n$, where $n$ is the number of qubits.  If $\cR = I$,
then $\rho_{out} = p C\ket{\mbox{\bf 0}}\bra{\mbox{\bf 0}}C^\dagger + {1\over
N}I$, so that
\ben
\trace(\rho_{out}\sigma_z^{(1)}) &=& p\ \trace(C\ket{\mbox{\bf
 0}}\bra{\mbox{\bf 0}}C^\dagger \sigma_z^{(1)}).
\een
If the excess probability $p$ of the ground state $\ket{\mbox{\bf 0}}$
is larger than the smallest detectable signal, we are able to
determine whether the output of a standard algorithm is $0$ or $1$ by
learning whether the measurement yields a negative or a positive
value.  To achieve sufficient confidence in the answer or to learn
more about the average answer, the bulk computation is repeated
several times. Confidence $c$ in the answer of a standard
algorithm at a signal to noise ratio of $\mbox{SNR}$ per experiment
requires $\sim \log(1/c)/\mbox{SNR}^2$ experiments.

Prior to the present work, there were two approaches to implementing
an effective pure state preparation procedure. These approaches may be
classified as {\em spatial averaging\/} and {\em logical labeling\/}.
Spatial averaging was introduced and implemented by Cory et
al.~\cite{cory:qc1996a,cory:qc1997a}. In general, spatial averaging
involves partitioning the ensemble of quantum computers into a number
of subensembles and applying a different unitary operator to each of
them. Given enough subensembles and proper choices of unitary
operators, the average density matrix over the whole ensemble can be
transformed into an effective pure state. This procedure requires
methods for distinguishing between quantum computers in the
ensemble. In NMR this can be accomplished by using well-known gradient
pulse methods to address individual cells in a bulk sample. The cells
in the implementation of Cory et al. are two dimensional slices of
constant magnetic fields defined by a transient gradient.  The logical
labeling technique of Gershenfeld and
Chuang~\cite{chuang:qc1996c,chuang:qc1997a} is fundamentally
different; it avoids the use of explicit subensembles by exploiting
ancillary qubits as labels. An initial unitary transformation is
applied which redistributes the states in such a way that conditional
on the state of the labels, an effective pure state is obtained in the
qubits to be used for computation. Gershenfeld and Chuang demonstrated
that this can be done efficiently in the high temperature limit for
non-interacting qubits, where $\rho_{0}$ can be expressed as a small
deviation from ${1\over N}I$.

Here, we consider a new and different technique: {\em Temporal
averaging}. Rather than attempting to guarantee an effective pure
state in a single experiment, this method uses several experiments
with different preparation steps chosen either systematically or
randomly. The measurements from each experiment are averaged to give
the final answer. The preparation steps are chosen such that the
average of the prepared input states is an effective pure state.  The
advantages of this method are that no ancillary qubits are needed, it
can be implemented to work at any temperature and it is not necessary
to distinguish subensembles of quantum computers.  In the high
temperature regime it can be implemented efficiently without any loss
of signal, and in general, the signal to noise ratios are sufficiently
well behaved to permit efficient determination of the desired answer
to any given level of confidence.

We will describe several temporal averaging methods and discuss their
properties. Temporal averaging methods can be loosely categorized into high
temperature and low temperature methods. The high temperature methods tend to
be simpler and are the most efficient for NMR quantum computations involving
small numbers of qubits. Three such methods will be described: Exhaustive
averaging, labeled flip\&swap and randomized flip\&swap. Labeled flip\&swap
uses a limited form of logical labeling to obtain the desired answer in two
experiments with only one ancilla, while randomized flip\&swap needs no
ancillas but may require additional experiments to overcome noise from the
randomization procedure. Flip\&swap methods rely on an inversion symmetry of
high temperature thermal states of non-interacting particles. Low temperature
methods do not require special assumptions on the initial state, but tend to
use more operations to implement. Two such methods are of interest,
randomization over a group and averaging by entanglement.  The first depends
on which unitary group is used.  We will show that there are groups which
yield good signal to noise behavior and which can be implemented in cubic
time. Averaging by entanglement has the advantage of requiring fewer
experiments, but necessitates discarding some of the qubits.  This method may
be useful if some of the qubits are discarded anyway for the purpose of
polarization enhancement by computational cooling, a family of techniques for
statically or dynamically increasing polarization of the ground state for a
subset of the available qubits.

The different temporal averaging methods are introduced and analyzed in the
following sections. We begin with a simple example borrowed from NMR, discuss
exhaustive averaging and the flip\&swap methods, show how randomized averaging
over a group can be used and give the method based on entanglement.  More
detailed descriptions of the algorithms and the mathematical analyses are in
the appendix. It is assumed that the reader is familiar with the basic
concepts of quantum
computation~\cite{ekert:qc1993a,yao:qc1993a,barenco:qc1996b} and nuclear
magnetic resonance~\cite{ernst:qc1994a}.

\section{NMR Example}

To illustrate the ideas on which temporal averaging is based, consider a two
qubit example from room temperature NMR with liquids.  The density matrix of
an AX system consisting of a proton and a carbon-13 nucleus in a 400MHz
spectrometer is approximately given by
\bea
	\rho = \frac{1}{4} 
		\left[\begin{array}{cccc}
			1 & 0 & 0 & 0
		\\	0 & 1 & 0 & 0
		\\	0 & 0 & 1 & 0
		\\	0 & 0 & 0 & 1
		\end{array}\right]
%
		+ 10^{-5} 
		\left[\begin{array}{cccc}
			1 & 0 & 0 & 0
		\\	0 & 0.6 & 0 & 0
		\\	0 & 0 & -0.6 & 0 
		\\	0 & 0 & 0 & -1
		\end{array}\right].
\eea
How to calculate these input states will be discussed below.  Because all
relevant observables are traceless, we focus our attention on the second
matrix, the {\em deviation density matrix}.  Suppose our goal is to perform
some computation $C$ on the ground state $\ket{00}\bra{00}$ and then to
observe $\sigma_z$ on the proton.  For this observation, the states
$\ket{01}\bra{01}$, $\ket{10}\bra{10}$ and $\ket{11}\bra{11}$ constitute
noise. To remove this noise we can exploit the fact that the computation and
the observation are linear in the input. We perform three experiments, each
with a different preparation step which permutes the undesirable input states,
and then average the output. The first experiment uses the unmodified input,
corresponding to preparation with $P_0 = I$. The second permutes
$\ket{01}\bra{01}\rightarrow\ket{10}\bra{10}\rightarrow\ket{11}\bra{11}
\rightarrow\ket{01}\bra{01}$ using the unitary transformation
\bea
P_1 &=& 
 \left[\begin{array}{cccc}
			1 & 0 & 0 & 0
		\\	0 & 0 & 1 & 0
		\\	0 & 0 & 0 & 1
		\\	0 & 1 & 0 & 0
		\end{array}\right].
\label{eq:pone}
\eea
This results in the input state
\ben
	\rho_1 = P_1 \rho P_1^\dagger = \frac{1}{4} 
		\left[\begin{array}{cccc}
			1 & 0 & 0 & 0
		\\	0 & 1 & 0 & 0
		\\	0 & 0 & 1 & 0
		\\	0 & 0 & 0 & 1
		\end{array}\right]
		+ 10^{-5} 
		\left[\begin{array}{cccc}
			1 & 0 & 0 & 0
		\\	0 & -0.6 & 0 & 0
		\\	0 & 0 & -1 & 0 
		\\	0 & 0 & 0 & 0.6
		\end{array}\right].
\een
The third preparation applies the inverse
permutation $P_2 = P_1^\dagger$ to
produce the input state
\ben
	\rho_2 = P_2 \rho P_2^\dagger \frac{1}{4} 
		\left[\begin{array}{cccc}
			1 & 0 & 0 & 0
		\\	0 & 1 & 0 & 0
		\\	0 & 0 & 1 & 0
		\\	0 & 0 & 0 & 1
		\end{array}\right]
		+ 10^{-5} 
		\left[\begin{array}{cccc}
			1 & 0 & 0 & 0
		\\	0 & -1 & 0 & 0
		\\	0 & 0 & 0.6 & 0 
		\\	0 & 0 & 0 & -0.6
		\end{array}\right].
\een
The average of the input density matrices
is then given by
\bea
\bar\rho &=& 
\frac{1}{3} \sum_i P_i \rho P_i^\dagger
\nonumber
\\ &=&
\frac{1}{4} 
		\left[\begin{array}{cccc}
			1 & 0 & 0 & 0
		\\	0 & 1 & 0 & 0
		\\	0 & 0 & 1 & 0
		\\	0 & 0 & 0 & 1
		\end{array}\right]
 + 10^{-5}
		\left[\begin{array}{cccc}
			1& 0 & 0 & 0
		\\	0 & -0.333 & 0 & 0
		\\	0 & 0 & -.333 & 0 
		\\	0 & 0 & 0 & -.333
		\end{array}\right]			\nonumber\\
 &=&
(\frac{1}{4}-0.333\cdot 10^{-5})
		\left[\begin{array}{cccc}
			1 & 0 & 0 & 0
		\\	0 & 1 & 0 & 0
		\\	0 & 0 & 1 & 0
		\\	0 & 0 & 0 & 1
		\end{array}\right]
 + 10^{-5}
		\left[\begin{array}{cccc}
			1.333& 0 & 0 & 0
		\\	0 & 0 & 0 & 0
		\\	0 & 0 & 0 & 0 
		\\	0 & 0 & 0 & 0
		\end{array}\right].
\label{eq:twoeps}
\eea
The average of the measurements of $\sigma_z^{(1)}$ after a computation
gives $\trace (C\bar\rho
C^{\dagger}\sigma_z^{(1)}) = 1.333 \cdot 10^{-5}\trace (C\ket{00}\bra{00}
C^{\dagger}\sigma_z^{(1)})$.  It can be seen that the contributions to
the measurements of the undesirable input states have been eliminated.
In NMR, $\sigma_z^{(1)}$ is measured by applying a radiofrequency
pulse to rotate the magnetization of the target spin into the plane
and observing the free induction decay as discussed
in~\cite{chuang:qc1997a}.

\section{Exhaustive Averaging}

The example of the previous section is an instance of {\em exhaustive
averaging\/}. For $n$ qubits, it involves cyclicly permuting the
non-ground states in $2^n-1$ different ways such that the average of
the prepared states is given by $(\rho_{00}-\bar p)\ket{\mbox{\bf
0}}\bra{\mbox{\bf 0}}+\bar p I$.  This method works for any initial state which
is diagonal in the computational basis states. Although the number of
experiments required grows exponentially, it is reasonable to consider
implementing it for small numbers of qubits.

\begin{figure}[htbp]
\begin{center}
\mbox{\psfig{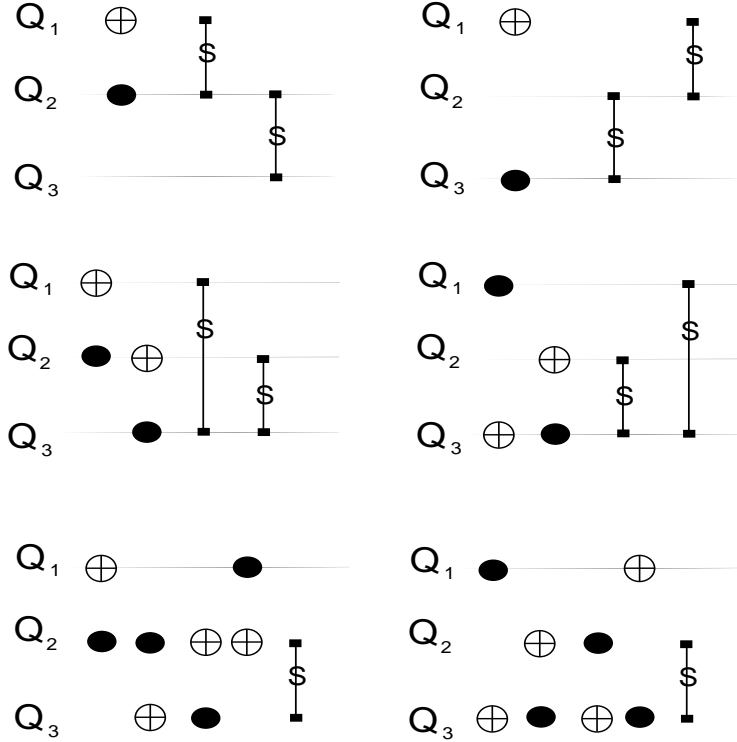}}
\end{center}
\caption{Networks required for state preparation when implementing exhaustive
averaging for three qubits using controlled-nots and swaps. The networks shown
perform the six non-identity cyclic permutations.  Seven experiments are
performed, one with no special preparation and six with the preparation
networks above.  $\oplus$ symbols denote the target qubits of the
controlled-not gates, and $\bullet$ symbols denote the control.}
\label{figure:ex-av-3}
\end{figure}

To design the quantum network for the preparation steps, one can
exploit the structure of the Galois field $\mbox{GF}(2^n)$. If the
non-ground initial states are labeled by elements of $\mbox{GF}(2^n)$,
multiplication by a non-zero element $x$ of this field implements one
of the cyclic permutations. Since multiplication can be implemented
with a reasonable (quadratic) number of controlled nots, each such $x$ yields a
preparation operator $P_x$.  Details are given in the appendix. The
seven networks needed to exhaustively average three qubits are in
Figure~\ref{figure:ex-av-3}.

The signal to noise ratio of exhaustive averaging is determined by the
sensitivity of each measurement, the excess probability in the ground state
and the number of experiments being performed.  If the initial density matrix
is $\rho = \sum_{i}\rho_{ii}\ket{i}\bra{i}$ with $0\leq i\leq 2^{n}-1$, then
the average density matrix over all experiments is given by $\bar\rho =
(\rho_{00}-\bar p)\ket{0}\bra{0} + \bar p I$, where $\bar p = {1\over
2^n-1}\sum_{i=1}^{2^n-1}\rho_{ii}$.  If the computation's output is
$x=\trace(C\ket{0}\bra{0}C^\dagger\sigma_z^{(1)})$, then the observed average
signal is $(\rho_{00}-\bar p)x$.  Given that the variance of the noise in each
measurement is $s^2$, the standard deviation of the noise in the average is
$s/\sqrt{2^n-1}$, which gives an overall signal to noise ratio of
${\sqrt{2^n-1}(\rho_{00}-\bar p)x/ s}$.  Typically, the density matrix will
describe a high temperature, polarized system of non-interacting spins, in
which case $\rho_{00} \approx n\delta/2^n$, where $\delta$ is the single spin
polarization (see Section~\ref{section:flip&swap}). It is also convenient to
define $\mbox{SNR}_1 = \delta/s$ as the signal to noise ratio from a single
spin measurement, such that we may express the signal to noise ratio of
exhaustive averaging as
\bea
	\mbox{SNR} &\geq& \frac{n}{2^n} \, \sqrt{2^n-1} |x| \mbox{SNR}_1
\,.
\eea
This argument assumes no bias in the individual measurements.  To
ensure that exhaustive averaging works correctly for standard quantum
algorithms, the bias must be small compared to $(\rho_{00}-\bar
p)/2^n$.

\section{Flip\&Swap}
\label{section:flip&swap}

{\em Flip{\sl\&}Swap\/} is a method which exploits special
properties of the high temperature thermal state for
non-interacting particles to create an effective pure
state with few experiments. 
If the internal Hamlitonian of a collection of $n$ qubits
is given by $\cH$, then the thermal state
is given by
\ben
\rho &=& \frac{e^{-\beta {\cal H}}}{\cal Z},
\een
where $\beta = 1/k_BT$ is the usual Boltzmann factor
and $1/{\cal Z}$ is the partition function normalization factor.
At high temperatures, a good approximation is to take
\ben
	\rho_{in} \approx {1\over N}(I - \beta {\cH})
,
\een
where $N = 2^n$ and we have defined energies so that $\trace\cH = 0$.

Consider the case where the Hamiltonian for the qubits is that of
non-interacting distinguishable particles with energy eigenstates
$\ket{0}$ and $\ket{1}$ and energies $-e_i$ and $+e_i$, respectively,
for the $i$'th qubit. This is a good approximation for many spin
systems in NMR, provided that the coupling constants are small
compared to the chemical shift differences between the different
spins. In this case, the energy eigenstates are close to the standard
computational basis states and the energy shifts due to coupling are
small compared to the Larmor frequencies.  The probability of the
state $\ket{b}$ for bit string $b=b_0b_1\ldots b_{n-1}$ is given by
\ben
\rho_{bb} &=& \prod_{i=0}^{n-1}{1\over 2}(1+ (-1)^{b_i}\delta_i)
\een
with
\ben
{1\over 2}(1+\delta_i) &=&
  {e^{\beta e_i}\over e^{\beta e_i} + e^{-\beta e_i}}.
\een
To first order, $\delta_i\approx \beta e_i$ is the
polarization of the $i$'th qubit.
Thus we can write
\ben
\rho_{bb} &=& {1\over N}(1+\sum_{i=0}^{n-1}(-1)^{b_i}\delta_i),
\een
where this first order approximation is valid as long as
$\delta_t \defeq \sum_{i=0}^{n-1}\delta_i\ll 1$.

\smallskip

Given the linear approximation to $\rho_{bb}$, it can be seen that if
$\bar b = (1-b_0)(1-b_1)\ldots (1-b_{n-1})$ is obtained from $b$ by
flipping each bit, then $\rho_{\bar b\bar b} + \rho_{bb} = {2\over
N}$.  Thus to obtain an unbiased, uniform input from two experiments,
it suffices to perform one experiment with no preparation
step and one with all the qubits
flipped in the preparation step, averaging the results.
However, this eliminates effective polarization in the
ground state as well as all the other states.

To retain
the ground state polarization we can perform two
experiments. In the first, the thermal input state is used
without modification by
applying preparation operator $P_0 = I$. In the
second, the preparation $P_1$ consists of
first inverting each qubit by applying $\sigma_x$
and then swapping the ground state $\ket{\mbox{\bf 0}}$
with the state $\ket{\mbox{\bf 1}}$ (all qubits in state $\ket{1}$).
The average of the two prepared states
is given by 
\bea
\rho_s &=&
 {1\over N} (I + \delta_t(
   \ket{\mbox{\bf 0}}\bra{\mbox{\bf 0}} - \ket{\mbox{\bf 1}}\bra{\mbox{\bf
	 1}})). 
\label{equation:def_rho_s}
\eea
There are two methods for eliminating the remaining polarization in
$\ket{\mbox{\bf 1}}$.  The first, {\em randomized\/} flip\&swap, uses
randomization to average this polarization over all non-ground
states. The second, {\em labeled\/} flip\&swap, uses one of the qubits
as a logical label, following the method
of~\cite{chuang:qc1996c,chuang:qc1997a}

The simplest randomization method involves first selecting a random
non-ground state $\ket{b}$ and applying a unitary operation $R$ which
maps $\ket{\mbox{\bf 1}}\rightarrow\ket{b}$ and leaves the ground
state unchanged. Both preparation steps are modified by adding this
unitary operation after the flip\&swap and before the computation.
To improve the signal to noise ratio, the whole procedure
can be repeated several times.
$R$ can be implemented efficiently using at most $n-1$
controlled-nots.

\begin{figure}[htbp]
\begin{center}
\mbox{\psfig{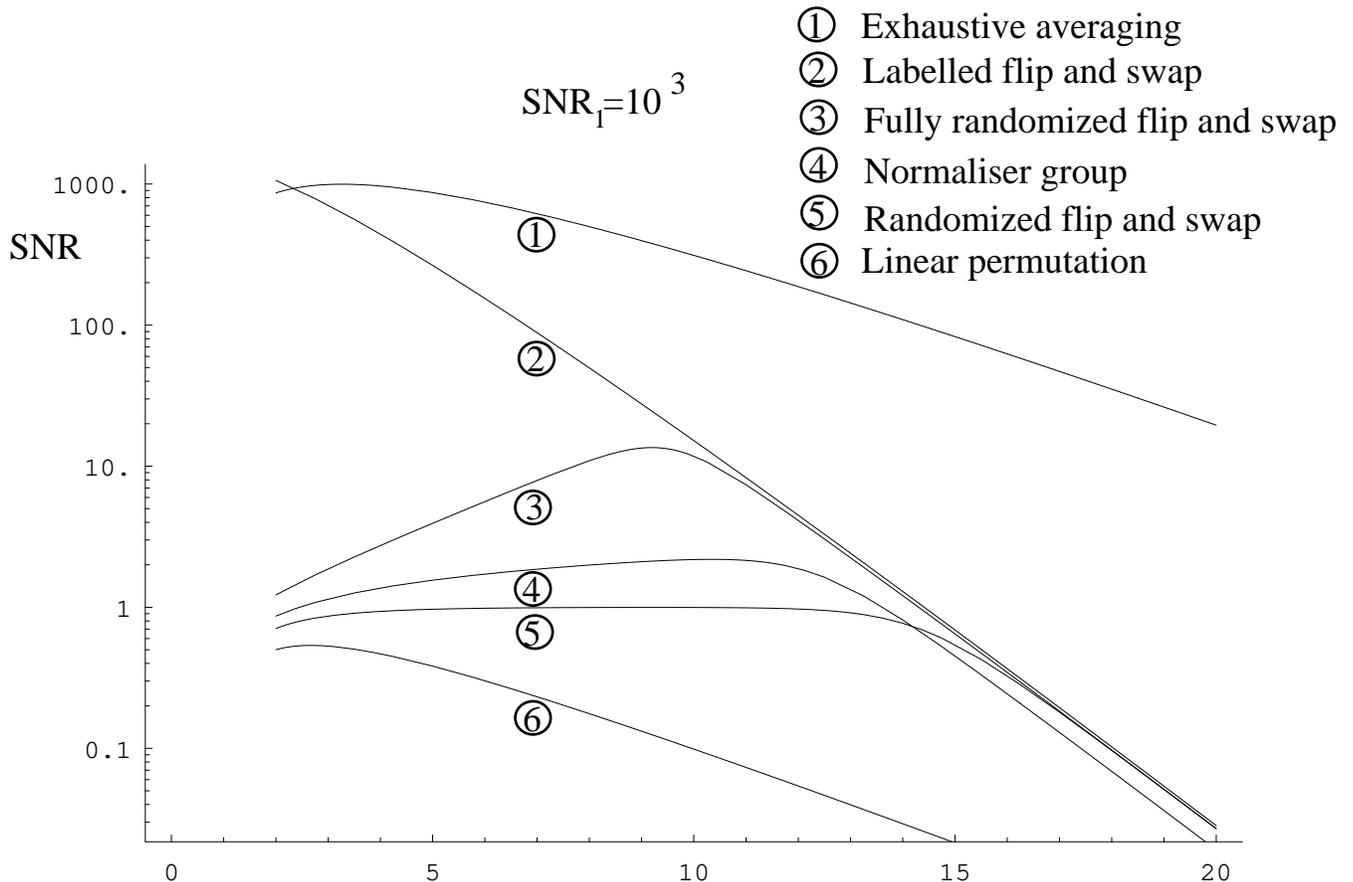}}
\end{center}
\caption{Graphs of lower bounds on the signal to noise ratio for the
different averaging methods for two or more identical non-interacting
qubits at high temperature, and $|x|=1$. The bounds hold for a one qubit
signal to noise ratio of $10^3$. The signal to noise ratios are for
one experiment in the case of randomization over a group, two
in the case of flip\&swap and $2^n-1$ in the case
of exhaustive averaging. The noise is due both to
experimental sensitivity and contributions from randomization (except
for labeled flip\&swap and exhaustive averaging,
which involve no randomization). Repeating
the experiments $k$ times with independent random choices increases
the signal to noise ratios by a factor of $k^{1/2}$.}
\label{figure:snr-comp}
\end{figure}

The signal to noise ratio for a randomized flip\&swap now depends not
only on the initial polarization of the ground state, the computation
and the sensitivity of the measurements, but also on the contribution
to the variance from the random choice of $R$.  The detailed
calculations of this variance will be given in the appendix. If all
the polarizations $\delta_i$ are the same, $\delta_i = \delta$, we
define $\mbox{SNR}_1 = \delta/s$ ( the signal to noise ratio for a
measurement $\sigma_z^{(1)}$ of the thermal state), and a lower bound
on the signal to noise ratio is given by 
\bea
	\mbox{SNR} &\geq&
	{n\over 2^n} {|x| \mbox{SNR}_1\over \sqrt{1/2 +
	n^2\mbox{SNR}_1^2/(2^n(2^n-2))}}.  
\eea
Graphs of the behavior of the signal to noise ratio of this and the
other methods are given in Figure~\ref{figure:snr-comp}. For small
$n$, the limited number of possible random choices results in a
significant reduction in the signal to noise ratio.  However, a
reasonable number of repetitions of the experiment can still reliably
determine any bias in $x$, if $x$ is not too small. An improvement in
the $\mbox{SNR}$ can also be obtained by using randomization over the
normalizer group as discussed in
Section~\ref{section:rand-group}. This is called {\em fully
randomized\/} flip\&swap.

\begin{figure}[htbp]
\begin{center}
\mbox{\psfig{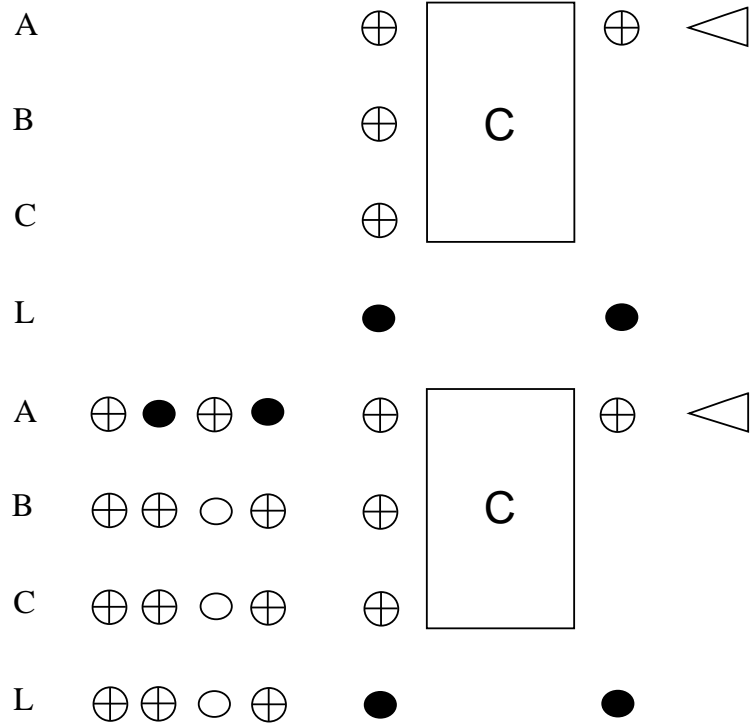}}
\end{center}
\caption{Quantum networks for the two experiments to implement
labeled flip\&swap for three computational qubits. The readout
operation on qubit A is shown explicitly as a triangle. Filled
circles denote conditioning on $|1>$, while unfilled
circles denote conditioning on $|0>$. }
\label{figure:lab-f&s}
\end{figure}

Labeled flip\&swap requires $n+1$ qubits and applies the flip\&swap operation
to all them.  Instead of removing the polarization in $\ket{\mbox{\bf 1}}$ by
averaging, it is exploited by using the $n+1$'th qubit as a label similar to
the methods introduced in~\cite{chuang:qc1997a}.  This method was discovered
independently by D. Leung.  Conditionally on the $n+1$'th qubit being in state
$\ket{0}$, the first $n$ qubits are in an effective pure state with excess
probability in $\ket{\mbox{\bf 0}}$.  Conditionally on the $n+1$'th qubit
being in state $\ket{1}$, the first $n$ qubits are in an effective pure state,
but with a deficiency in $\ket{\mbox{\bf 1}}$.  Both experiments' preparation
steps must be followed by an operation which conditionally on the $n+1$'th
qubit flips all the other qubits to turn the conditional deficiency in
$\ket{\mbox{\bf 1}}$ into one in $\ket{\mbox{\bf 0}}$. After the computation
is complete, the deficiency can be turned into an effective excess by
conditionally reversing the sign of the answer. The full network for $n=3$ is
given in Figure~\ref{figure:lab-f&s}.

The signal to noise ratio for labeled flip\&swap is given by
\bea
\mbox{SNR} &=&
  { \sqrt{2}(n+1) |x| \mbox{SNR}_1 \over 2^n}.
\eea

\section{Randomization over Groups}
\label{section:rand-group}

Exhaustive averaging is useful for small numbers of qubits and
flip\&swap works for nearly non-interacting qubits
at high temperatures.  If the number of qubits and the
polarization satisfy $n\delta \sim 1$ or if the initial
state does not have approximate inversion symmetry, it is
necessary to consider other methods which are both reasonably
efficient and can be applied to arbitrary initial states.
Randomization based on groups of unitary operators has this
property.

In general, randomization involves choosing a preparation operator $P$
according to a predetermined probability distribution.  To ensure that the
expected value of the measurement represents the output of the
computation on an effective pure state, we require that $\Exp_P(P\rho
P^\dagger) = \bar\rho$ is an effective pure state.
The methods to be discussed satisfy
that
\ben
\bar\rho &=& (\rho_{00}-\bar p)\ket{0}\bra{0}
              + \bar p I,
\een
with $\bar p ={1\over N-1}\sum_{i\geq 1}\rho_{ii}$.  It is desirable
that the initial state $\rho$ has excess probability in the ground
state. If possible, the true initial state should be transformed by a
unitary transformation which guarantees that the maximum probability
state is the ground state, and that the density matrix is diagonal in
the computational basis. (For nearly uniform mixtures of states and
high sensitivity, it may be more efficient to have a sufficiently
large deficiency in the ground state.)

Let $\sigma = C^\dagger \sigma_{z}^{(1)} C$,
so that $x = \trace(\ket{0}\bra{0}\sigma) = \sigma_{00}$.
A single experiment with randomized preparation
yields the measurement $r(P) = \trace(P\rho P^\dagger\sigma)$
with variance $s^2$;
the expectation of $r(P)$ is given
by $\bar r = (\rho_{00}-\bar p) x$.
The signal to noise ratio for a single run of the computation is
determined by comparing the variance $\bar v$ of $r(P)$ to $\bar r^2$.
Thus, the signal to noise ratio is
\be
	\mbox{SNR}(P,C,\rho) = {|\bar r|\over \sqrt{s^2+ \bar v}}.
\ee
If, for example, we wish to learn the expectation of $\bar r$ to
within $\bar r (1\pm\epsilon)$, the number of experiments required to
achieve confidence $c$ is proportional to
$\log(1/c)/(\epsilon^2\mbox{SNR}(P,C,\rho)^2)$ in the Gaussian
regime. Due to the large number of choices in the randomization it is
reasonable to expect that this regime applies even for one experiment.
If this is were not the case, the average would need to be inferred
by techniques robust against outliers. If we are only
interested in learning the sign of $\bar r$ with confidence $c$, this
can be done with $\sim\log(1/c)/\mbox{SNR}(P,C,\rho)^2$ experiments,
regardless of the actual distribution. One method is to use the sign
of the median of the $k_1$ averages of the results from
$k_2\defeq\max(1, 4/\mbox{SNR}(P,C\rho)^2)$ independent experiments.
Because the probability of the event that the average of $k_2$
experiments has the wrong sign is bounded by $1/4$, the probability of
failure is $\leq e^{-O(k_1)}$.  The constant in the exponent can be
obtained from the Chernoff-Hoeffding bounds~\cite{alon:qc1992a} for
the probablity of having more than $1/2$ heads in $k_1$ flips of a
biased coin with the probability of head given by $1/4$.

To compute the variance
of $r(P)$, define
\ben
\check\rho &\defeq&
	\rho-\Exp_P P\rho P^\dagger\nonumber\\
    &=& \rho - \bar p I - (\rho_{00}-\bar p)\ket{0}\bra{0}.
\een
Then
\ben
\bar v &=& \Exp_P\trace(P\check\rho P^\dagger\sigma)^2\nonumber\\
       &=& \Exp_P\trace((P\check\rho P^\dagger\tensor P\check\rho P^\dagger)
           (\sigma\tensor\sigma))\nonumber\\
       &=&\trace(\Exp_P (P\check\rho P^\dagger\tensor P\check\rho P^\dagger)
           (\sigma\tensor\sigma)).
\een
Thus to ensure that $\bar r$ is as desired and to compute $\bar v$, we
first verify that $\Exp_P (P\check\rho P^\dagger) = 0$ and then
compute $\Exp_P (P\check\rho P^\dagger\tensor P\check\rho P^\dagger)$.

In the algorithms described below, $P$ is a random product of
operators, each chosen uniformly from various groups of unitary operators.
The desired expectations can often be computed in closed form if $P$
is a random element of a unitary group $G$.  
For this purpose, it is convenient to use the representations of $G$ defined
by $\pi_1(P)(A) = PAP^\dagger$ and $\pi_2(P)(A\tensor B) = PAP^\dagger\tensor
PBP^\dagger$, where $\pi_2(P)$ is linearly extended to all four-tensors.  Both
$\pi_1$ and $\pi_2$ are unitary representations of $G$ for the usual inner
product of operators and four-tensors: $\langle A, B\rangle = \trace(AB)$ and
$\langle A\tensor B, C\tensor D\rangle = \trace(AC)\trace(CD)$, with the
latter inner product extended bilinearly to all four-tensors.  Using this
representation, for $P$ sampled uniformly from $G$, it follows that the
expectations can be obtained by projecting $\rho$ and $\rho\tensor\rho$ onto
the trivial eigenspaces of $\pi_1$ and $\pi_2$.  Specifically, let $\Pi_1$ and
$\Pi_2$ be the projection superoperators onto the space of all $A$ such that
$\pi_1(P)A = A$ and onto the space of all $B$ such that $\pi_2(P)B = B$,
respectively. Then $\Exp_{P\in G} \pi_1(P)A^\dagger = \Pi_1 A$ and $\Exp_{P\in
G} \pi_2(P)B = \Pi_2 B$.  We use this to calculate variances resulting from
averaging over four groups, below.

\subsection{Diagonal Groups}
\label{section:diagonal}

If the initial density matrix is not diagonal and it is not feasible
to perform the unitary transformation which makes it diagonal in the
computational basis, one can use randomization over a diagonal group
to reduce the effect of the offdiagonal entries.  Let $\cD$ be a group
of diagonal operators $S_f:\ket{j}\rightarrow
i^{f(j)}\ket{j}$, with $f(j)\in \{0,1,2,3\}$.  To ensure
sufficiently small trivial eigenspaces for the representations $\pi_1$
and $\pi_2$, we require that the following
phase independence condition holds: If $f(j)-f(k)+f(l)-f(m) = 0\mod(4)$ for
all $f$, then $j=k$ and $l=m$ or $j=m$ and $k=l$. We call a group with
this property a {\em diagonal group\/}. Randomization
over $\cD$ is accomplished by choosing a member of $\cD$
uniformly and applying it to the initial state.
Although the expectation of the randomized density matrix
is not yet an effective pure state, it does reduce
the off-diagonal contributions to the expectation and the
variance. For example,
\ben
\Exp_{P\in \cD} P\rho P^\dagger &=& \sum_{i=0}^{N-1}\rho_{ii}\ket{i}\bra{i}.
\een
To obtain an effective pure state, additional randomization
steps are required. 
The expectations needed for computing variances are calculated in the
appendix. An efficiently implementable
diagonal group $\cD$ can be obtained as a subgroup of
the normalizer group introduced below.

\subsection{ Two-transitive Permutation Groups}

Let $\cT$ be a two-transitive group of permutations acting on the set of states
$\ket{1},\ldots,\ket{N-1}$. By definition, for every $i\not=j$ and $k\not=l$,
there is a permutation $\pi\in \cT$ such that $\pi(i) = k$ and $\pi(j) = l$.
Then
\ben
\Exp_{P_1\in \cD, P_2\in \cT} P_2P_1 \rho P_1^\dagger P_2^\dagger &=&
  (\rho_{00}-\bar p)\ket{0}\bra{0} + \bar p I,
\een
which is the desired effective pure state. An effective
pure state would be obtained on average even with a one-transitive
group, such as the cyclic permutations used for exhaustive
averaging. However, the variance for one-transitive groups
can be quite large and two-transitivity helps in
reducing it. 

To give the upper bound on $\bar v$  for randomization
with $\cD$ and $\cT$, define
\ben
\check\rho_d &\defeq& \sum_{i\geq 1} \check\rho_{ii}\ket{i}\bra{i},
\een
Then
\ben
\bar v &\leq& \trace(\check\rho_d^2) +{1\over N-2}\trace(\check\rho^2).
\een
The derivation of this inequality is in the appendix.
In the high temperature regime, this implies a signal
to noise ratio of at least
\bea
	\mbox{SNR} &\geq&  {n\over 2^n}{|x|\mbox{SNR}_1\over \sqrt{1 +
	n\mbox{SNR}_1^2/(2^n-2)}}.  
\eea

Efficiently implementable two-transitive permutation groups
can be obtained from the normalizer group.

\subsection{The Normalizer Group}
\label{section:normalizer-group}

The normalizer group $\cN$, more specifically, the normalizer of
the error group, consists of all unitary operations $U$ which satisfy
that for any tensor product of Pauli operators $\sigma$, $U\sigma
U^\dagger$ is also a tensor product of Pauli operators (up to a phase
factor).  If the Pauli operators are labeled by $\sigma_{00}\defeq I,
\sigma_{01}\defeq\sigma_z,\sigma_{10}\defeq\sigma_x,
\sigma_{11}\defeq\sigma_y$ and, for example, $\sigma_{101101} =
\sigma_{10}\tensor\sigma_{11}\tensor\sigma_{01}$, then the elements of the
normalizer group are characterized by $U\sigma_b U^\dagger =
(-1)^{\langle x,b\rangle} i^{f(b,L)}\sigma_{Lb}$, where $x$ is an
arbitrary bit vector, $\langle x,b\rangle$ denotes the inner product
modulo $2$ of bit vectors and $L$ is an arbitrary invertible (modulo
2) 0-1 matrix which satisfies $L^T M L = M$, where $Mb$ is the bit
vector obtained from $b$ by swapping adjacent bits belonging to the
same factor. The exponent
$f(b,L)$ depends only on $b$ and $L$; its values
are not needed for the present analyses.
The group of matrices $L$ with this property acts
transitively on non-zero bit vectors. 
The normalizer group yields several subgroups useful for
randomization.

\medskip

\noindent{\bf Linear Phase Shifts.}

The group $\cD$ generated by controlled-sign flips and the operator $S =
\scriptstyle\left(\begin{array}{cc}1&0\\0&i\end{array}\right)$ acting on any
qubit consists of diagonal operators with action $\ket{k}\rightarrow
i^{\langle x, k\rangle} (-1)^{\langle k, Bk\rangle}\ket{k}$, where $x$ is a
vector with entries in $\{0,1,2,3\}$ and $B$ is an arbitrary $n$ by $n$ 0-1
matrix.  To check that the phase independence condition (Section~\ref{section:diagonal}) holds, suppose that
for all $x$ and $B = yz^T$,
\ben
x^T(k-l+m-n) + 2(k^T y z^T k - l^Ty z^Tl + m^T y z^T m - n^T y z^T n)
   &=& 0\mod(4).\\
\een
This implies that $k-l+m-n = 0\mod(4)$. If $k=m$, then $l=n=k$, since
$k,l,m$ and $n$ are all 0-1 vectors.  If not, without loss of
generality, suppose that $k\not=0$.  To derive a contradiction,
suppose also that $k\not=l$ and $k\not=n$. If $k$ is not in the span
(modulo two) of $l,m$ and $n$, then there exists $z$ orthogonal
(modulo two) to $l,m$ and $n$, but not $k$, which contradicts the
equality above. Thus $k$ must be in the span of $l,m$ and $n$.  If $k$
is not in the span of two of them, say $l$ and $m$, then there exists
$z$ orthogonal to $l$ and $m$ but not $k$ and a $y$ orthogonal to $n$
but not $k$.  Again, we find that the equality cannot hold.  Thus it
must be the case that $k=l+m\mod(2)$, $k=l+n\mod(2)$ and $k=m+n\mod
2$. This implies that $m=n=l$ and $k=0$. Thus the desired independence
condition holds.

\medskip

\noindent{\bf Linear Cyclic Permutations.}

A group $\cS$ acting cyclicly on the set $\ket{1},\ldots,\ket{n}$ is
obtained by representing the field $\mbox{GF}(2^n)$ as a vector space over
$\mbox{GF}(2)$ with elements represented by bit strings of length $n$ in some
basis. Multiplication by non-zero elements of $\mbox{GF}(2^n)$ defines
a cyclic subgroup of $\cL$ of order $2^n-1$.

\medskip

\noindent{\bf Linear Permutations.}  

The group $\cT$ of linear permutations is generated by the
controlled-not operations. The group consists of the unitary $U$'s
which satisfy $U\ket{b} = \ket{Lb}$, where $L$ is an invertible
(modulo two) 0-1 matrix. The group acts two-transitively on the set
$\ket{1},\ldots,\ket{N}$.

\subsection{Conditional Normalizer Group}

Randomization over the normalizer group is as effective for variance
reduction as is randomization over the unitary group. The main
difficulty is that the normalizer group does not fix $\ket{0}$. This
can be remedied by alternating randomization with $\cT$ and with the
conditional normalizer group $\cN_1$ which acts on the first $n-1$
qubits given that the last one is in state $\ket{1}$.

The first step in the procedure is to randomize with $\cD$ (if needed)
and $\cT$.  Each following step involves randomizing with $\cN_1$ and
then with $\cT$.  The total number of steps determines how effective
the randomization is. The procedure is designed such that the
expectation of the resulting density matrix is the desired effective
pure state after every step.
The variance $\bar v_{k+1}$ after the $k$'th step can be estimated
by (see the appendix):
\ben
\bar v_{k+1}&\leq&
    \lambda^k{N\over N-2}\trace(\check\rho_d^2)
        +{1\over N-2}\check\rho^2,
\een
with $\lambda \defeq {e^{1\over N+2}/2}$.
In the high temperature regime this implies a signal
to noise ratio of
\bea
	\mbox{SNR} &\geq& 
	  {n\over 2^n}{|x|\mbox{SNR}_1\over 
	  \sqrt{1+2n\mbox{SNR}_1^2/(2^n(2^{n}-1))}},
\eea
where $k$ was chosen such that $\lambda^k\leq {1/2(2^n+2)}$.

\section{Effective Pure States by Entanglement}

The temporal randomization methods discussed above are useful when the
device is qubit limited, in the sense that it is difficult to access
additional qubits. It is important to realize that
ancillary qubits involved only in preparation and postprocessing
generally do not need to have long decoherence or relaxation
times. For example, if they are used only in the preparation phase,
their quantum coherence does not need to be maintained in computation or
readout. If such ancillas are available, they can and should be used
to simplify the effective pure state preparation.  Interestingly, if
an additional $n$ qubits are available, it is possible to prepare a
nearly perfect effective pure state for any diagonal initial state by
exploiting entanglement.

Here is an explicit algorithm which results in an effective pure state
on the first $n$ qubits given $2n$ qubits. The basic idea is to map
the computational basis states other than the ground state on the
first $n$ qubits to nearly maximally entangled states.  Write a
computational basis state on the $2n$ qubits as $\ket{a}\ket{b}$,
where $a$ and $b$ are length $n$ bit vectors.  Let $x$ be a generator
of the multiplicative group of non-zero elements of $\mbox{GF}(2^n)$.
The desired unitary transformation is the composition of the maps
\ben
P_1&:&\ket{a}\ket{b}\rightarrow\sum_{c}(-1)^{\langle b,c\rangle}\ket{a}\ket{c},\\
P_2&:&\ket{a}\ket{b}\rightarrow\ket{a x^b}\ket{b},
\een
where $b$ is interpreted as a bit vector in the first exponent and as
a binary number in the second.  Consider
the reduced density matrix $\varrho_{ab}$ on the first $n$ qubits
derived from the state $P_2P_1\ket{a}\ket{b}$. If $a\not=0$,
\ben
\varrho_{ab} &=& {1\over N}(I-\ket{0}\bra{0} + \ket{a}\bra{a})\\
\varrho_{0b} &=& \ket{0}\bra{0}.
\een
This is nearly an effective pure state. If $\rho$ is
the reduced initial density matrix on the first $n$ qubits and
$\rho$ is diagonal,
then after applying $P_2P_1$, the reduced density matrix is
\ben
{N-1\over N}((\rho_{00}-\bar p)\ket{0}\bra{0}
 + \bar p I) + {1\over N}\rho.
\een
The deviation from the effective pure state is sufficiently small
to be of no concern in most cases. 

Entanglement can be exploited even if less than $n$ additional
bits are available. In fact, essentially the same algorithm
works. However, the deviation from an effective pure state
becomes larger and residual bias must be removed by
another technique such as randomization.
In general, if ancillary qubits are available, the effectiveness of
averaging methods can be improved.  For example, we can randomize the
states $\ket{a}\ket{b}$ with $a\not=0$ with the subgroup $\cL_m$ of
the group of linear permutations which preserves the subspace
$\{\ket{0}\ket{b}\}$.  If this does not reduce the variance enough, a
version of the conditional normalizer randomization method can be
used, where $\cL_m$ is used instead of the full group of linear
permutations.

Ancillary qubits are likely to be available whenever a computational
cooling method is used to increase the probability of the
ground state in some of the available qubits. Computational cooling
uses ancillas and in-place operations to transfer heat from
the computational qubits to the ancillas. The simplest such
methods are based on decoding a classical error-correcting
code in-place and exploiting the fact that the thermal
state is equivalent to a noisy ground state.

\section{Experimental Evidence}

The temporal randomization methods can find immediate
application in NMR quantum computation, even with simple molecules, as we
demonstrate with the following experimental results utilizing
exhaustive averaging to extract an effective pure state
from a two spin system.

\begin{figure}[htbp]
\begin{center}
\mbox{\psfig{file=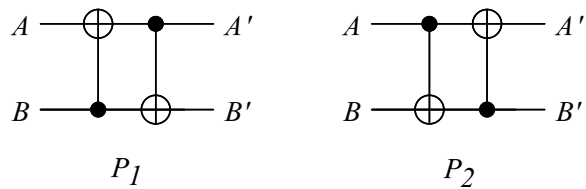,width=3.25in}}
\end{center}
\caption{Quantum circuit implementation of the permutations $P_1$ and $P_2$.}
\label{fig:permcirc}
\end{figure}

Using a model two spin system, we prepared an effective state similar
to that of Eq.(\ref{eq:twoeps}) from a thermal state.  This was done
by implementing the quantum circuits shown in Fig.~\ref{fig:permcirc}
to perform the permutation of Eq.(\ref{eq:pone}) and its inverse.

\begin{figure}[htbp]
\begin{center}
\mbox{\psfig{file=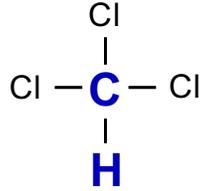,width=1in}}
\end{center}
\caption{Molecule of chloroform: the two active spins in
	this system are the $^{13}$C and the $^1$H.}
\label{fig:chloroform}
\end{figure}

The two-spin physical system used in these experiments was carbon-13 labeled
chloroform (Fig.~\ref{fig:chloroform}) supplied by Cambridge Isotope
Laboratories, Inc. (catalog no. CLM-262), and used without further
purification.  A 200 millimolar sample was prepared with d6-acetone as a
solvent, degassed, and flame sealed in a standard 5mm NMR sample tube, at the
U.C. Berkeley College of Chemistry.

Spectra were taken using Bruker AMX-400 (U.C. Berkeley) and DRX-500 (Los
Alamos) spectrometers using standard probes.  The resonance frequencies of the
two proton lines (in the DRX-500) were measured to be at $500.133921$ MHz and
$500.134136$ MHz, and the carbon lines were at $125.767534$ MHz and
$125.767749$ MHz, with errors of $\pm 1$ Hz.  The radiofrequency (RF)
excitation carrier (and probe) frequencies were set at the midpoints of these
peaks, so that the chemical shift evolution could be suppressed, leaving only
the $215$ Hz J-coupling between the two spins.  The $T_1$ and $T_2$ relaxation
times were measured using standard inversion-recovery and
Carr-Purcell-Meiboom-Gill pulse sequences.  For the proton, it was found that
$T_1 \approx 7$ sec, and $T_2 \approx 2$ sec, and for carbon, $T_1\approx 16$
sec, and $T_2 \approx 0.2$ sec.  The short carbon $T_2$ time is due to
coupling with the three quadrupolar chlorine nuclei, which shortens the
coherence time.  Nevertheless, these time scales were all much longer than
those of the operations applied, guaranteeing that we could implement quantum
transforms and observe quantum dynamics.

\begin{figure}[htbp]
\begin{center}
(A)~\mbox{\psfig{file=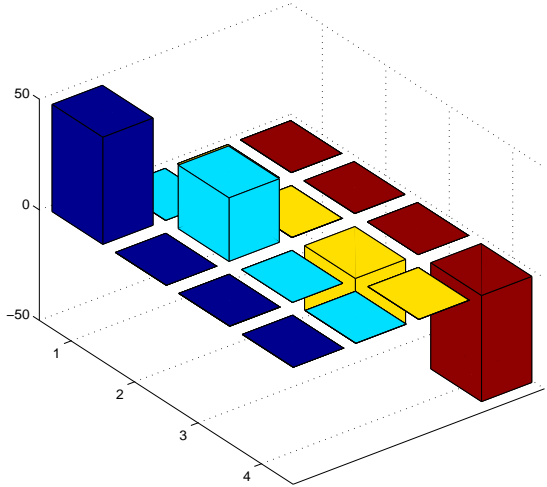,width=3.25in}}~~~~(B)~\mbox
	{\psfig{file=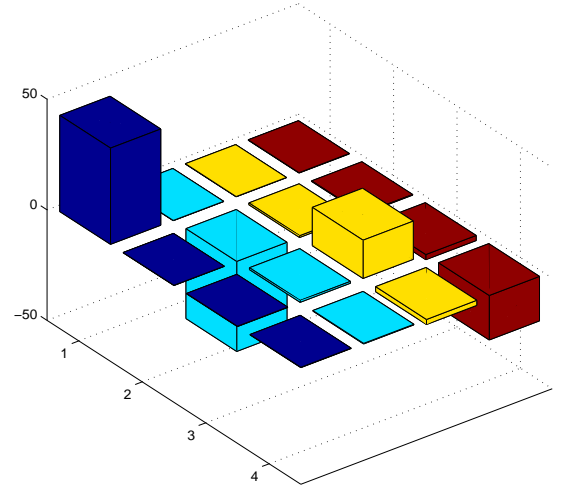,width=3.25in}}	\\
(C)~\mbox{\psfig{file=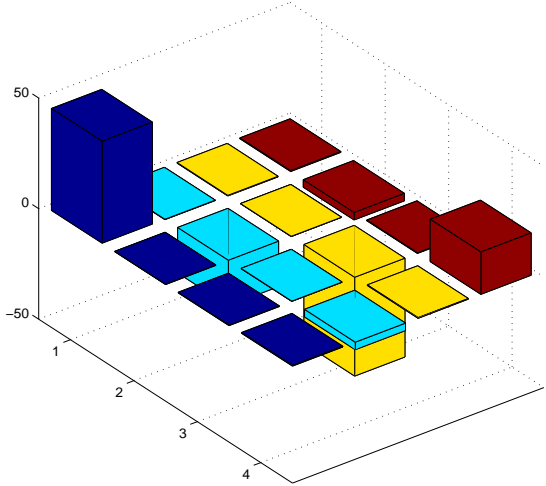,width=3.25in}}~~~~(D)~\mbox
	{\psfig{file=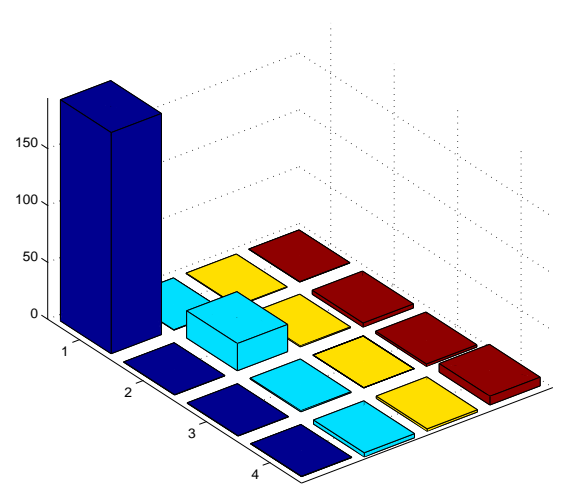,width=3.25in}}
\end{center}
\caption{Experimentally measured deviation density matrices for
	 (A) thermal state,
	 (B) state after $P_1$ operation,
	 (C) state after $P_2$ operation.
	 (D) Effective pure state (biased sum of the three).  Real components
	 only are shown; all imaginary components are small.}
\label{fig:denmats}
\end{figure}

We performed {\em quantum state tomography} to systematically obtain the final
quantum state; this procedure will be described in detail
elsewhere~\cite{chuang:qc1997b}.  In each tomography procedure, nine
experiments were performed, applying different pulses to measure all the
possible elements in the density matrix in a robust manner.  The resulting
deviation density matrix for the thermal state is shown in
Fig.~\ref{fig:denmats}A.  As expected, all the off-diagonal elements are
nearly zero, while the diagonal elements follow a pattern of $a+b$, $a-b$,
$-a+b$, and $-a-b$.  An error of about $5$\% was observed in the data, due
primarily to imperfect calibration of the $90^\circ$ pulse widths and
inhomogeneity of the magnetic field.

\begin{figure}[htbp]
\begin{center}
\mbox{\psfig{file=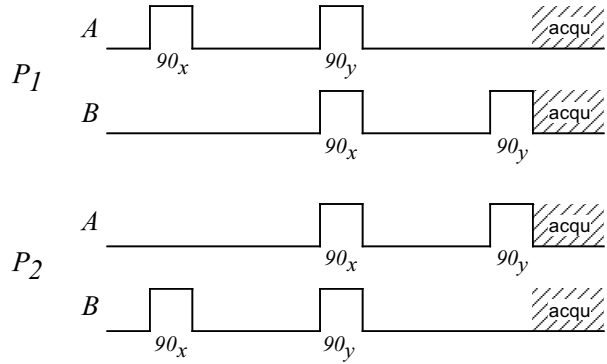,width=3.25in}}
\end{center}
\caption{NMR pulse program implementations of the permutations $P_1$ and
	$P_2$.  Each RF pulse was about $10$ microseconds long, and
	the time between the pulses was about $2.3$ milliseconds.}
\label{fig:permpulses}
\end{figure}

The two permutation quantum circuits were implemented using the pulse
programs shown in Fig.~\ref{fig:permpulses}.  Because of the absence of phase
correction steps in the controlled-not gates\cite{chuang:qc1997a}, the actual
transforms implemented were not exactly those of
Eq.(\ref{eq:pone}), but rather,
\bea
	\tilde P_1 &=& \left[\begin{array}{cccc}
			i & 0 & 0 & 0
		\\	0 & 0 & -1 & 0
		\\	0 & 0 & 0 & 1
		\\	0 & i & 0 & 0
		\end{array}\right]
\label{eq:ponetilde}
\\
	\tilde P_2 &=& \left[\begin{array}{cccc}
			i & 0 & 0 & 0
		\\	0 & 0 & 0 & 1
		\\	0 & -1 & 0 & 0
		\\	0 & 0 & i & 0
		\end{array}\right]
\label{eq:ptwotilde}
\,.
\eea
For the purposes of temporal randomization of an initially diagonal
density matrix, the phases of the transformations can be ignored.  We
obtained the density matrices shown in Fig.~\ref{fig:denmats}B-C from
these two transformations.  The effective pure state we obtained was
approximately
\be
	\bar\rho = 
		\left[\begin{array}{cccc}
			194 & \ep & \ep & \ep
		\\	\ep & 24 & \ep & \ep
		\\	\ep & \ep & \ep & \ep 
		\\	\ep & \ep & \ep & 8
		\end{array}\right]
	-
	57 I
\label{eq:twoepsexpt}
\,,
\ee
where $|\ep|<5.4$.  An error of $\pm 5$ was calculated, based on
analysis of the linewidth integration, least squares fitting used in
the tomography procedure, and standard error propagation.  This result
compares favorably with the result expected from Eq.(\ref{eq:twoeps}).
Further work has been done to use this state as an input into a
non-trivial computation; that work demonstrates the creation and
manipulation of effective pure states which are in superpositions, and
will be reported elsewhere\cite{chuang:qc1997b}.

\section{Conclusion}

We have described new techniques for creating effective pure states
which complement the logical labeling and spatial averaging techniques
previously discovered.  Our {\em temporal averaging} methods are
unique in their use of summation over experiments carried out at
different times and powerful by virtue of averaging over
transformations chosen systematically (in the case of labeled
flip\&swap) or randomly (for randomization over a transformation
group).  The choice of temporal averaging method in an experiment
depends on the number of qubits available, how many are required for
computation, the initial density matrix and the desired signal to
noise ratio. A summary of our recommendations based on the analyses in
this manuscript follows: For small numbers of qubits exhaustive
averaging can be used for any initial density matrix which is diagonal
in the computational basis. If the initial state is close to that of
non-interacting particles at high temperature, the flip\&swap
techniques can be used.  If a non-computational qubit is available,
then labeled flip\&swap is the simplest and most efficiently
implemented method. Asymptotically it requires a linear number of
quantum operations, and unless high signal to noise ratios are needed,
involves many fewer experiments than exhaustive averaging. In terms of
quantum operations, exhaustive averaging appears to be more efficient
up to at least four qubits.  The actual minimum number of qubits for
which labeled flip\&swap uses fewer quantum operations per experiment
then exhaustive averaging depends on the implementation and remains to
be determined. If every qubit is required for computation, then
randomized flip\&swap can be used at a cost of more quantum operations
per experiment. For large numbers of qubits where the high temperature
regime or the non-interacting assumption fails, randomization over a
group can be used. If ancillary qubits are available, randomization
can be combined with entanglement.  It remains to be seen whether this
situation will be encountered in practice.

Future theoretical work will investigate combinations of logical, spatial, and
temporal labeling techniques, and establish a connection between these
procedures and error-correction.  Experiments will also be performed to
demonstrate the different techniques
with large molecules and to explore their relative
merits in practice.


\noindent{\bf Acknowledgements.}  D. Leung independently discovered that
labeling and temporal averaging could be combined in the high temperature
regime by labeled flip\&swap.  Thanks to W. Zurek for encouraging our work on
NMR quantum computation. We profited from stimulating discussions with
J. Anglin and D. Divincenzo, and ackowledge M. Kubinec, P. Catasti and
S. Velupillai for experimental assistance. In particular we thank
S. Velupillai for suggesting the use of labeled chloroform in the experiment.  This work was performed
under the auspices of the U.S. Department of Energy under Contract
No. W-7405-ENG-36.


\appendix

\section{Calculations of Variance for Randomization over Groups}

The expectation and variance of the outcome of an experiment using
randomization over a group $G$ can be determined from the trivial
eigenspaces of the representations $\pi_1:\pi_1(U)(A) = UAU^\dagger$
and $\pi_2:\pi_2(U)(A\tensor B) = UAU^\dagger\tensor UBU^\dagger$. In
the next sections, these eigenspaces are determined and the
resulting variances estimated.  We begin with some calculations for
the diagonal groups.

\subsection{Diagonal Groups}

Let $\cD$ be a diagonal group as defined in Section~\ref{section:diagonal}.
This group is used to diagonalize the average density
matrix before randomizing with more powerful groups.
We compute the projections onto the trivial eigenspaces
of both representations $\pi_1$ and $\pi_2$.
\ben
\Exp_{P\in \cD} P\ket{i}\bra{j}P^\dagger &=& \delta_{i,j}\ket{i}\bra{i},
\een
and
\ben
\Exp_{P\in \cD} P\ket{i}\bra{i}P^\dagger\tensor P\ket{k}\bra{k}P^\dagger
  &=& \ket{i}\bra{i}\tensor \ket{k}
	\bra{k},
\\
\Exp_{P\in \cD} P\ket{i}\bra{k}P^\dagger\tensor P\ket{k}\bra{i}P^\dagger
  &=& \ket{i}\bra{k}\tensor \ket{k}
	\bra{i}.
\een
Other expectations of $\ket{i}\bra{j}\tensor\ket{k}\bra{l}$
are $0$. The projections of $\check\rho$ and $\check\rho\tensor\check\rho$
onto the trivial eigenspaces of $\pi_1$ and $\pi_2$ are therefore
given by
\ben
\Exp_{P\in \cD} P\check\rho P^\dagger &=& \sum_{i\geq 0}\check\rho_{ii}\ket{i}\bra{i},\\
\Exp_{P\in \cD} P\check\rho P^\dagger\tensor P\check\rho P^\dagger &=&
  \sum_{i,j\geq 0}\check\rho_{ii}\ket{i}\bra{i}\tensor \check\rho_{jj}\ket{j}\bra{j}
  +
  \sum_{i\not=j\geq 0}\check\rho_{ij}\ket{i}\bra{j}\tensor\check\rho_{ji}\ket{j}\bra{i}.\\
\een
Unfortunately, it is impossible to completely eliminate the
contributions of the off-diagonal elements of $\rho$ to the variance
by this method. As will be seen, to reduce the effect of these
contributions it is necessary to ensure that $\rho$ is approximately
diagonal by an initial unitary operation, or to design the algorithm
so that $\sigma$ is approximately diagonal (as will be the case if the
output of the algorithm is deterministic when given one of the
computational basis states for input). 

The calculations for the other groups to be presented below assume
that $\rho$ has already been randomized by a diagonal group. As a
result, only the subspaces spanned by $\ket{i}\bra{i}$ (for $\pi_1$)
and by $\ket{i}\bra{i}\tensor\ket{j}\bra{j}$ (for $\pi_2$) and
$\ket{i}\bra{j}\tensor\ket{j}\bra{i}$ will be considered
in our analysis.

\subsection{ Two-transitive Permutation Groups}

Let $\cT$ be a two-transitive permutation group which
fixes $\ket{0}$. 
It is straightforward to check that for $i\not=j$
\ben
\Exp_{P\in \cT} P\ket{i}\bra{i}P^\dagger &=& {1\over N-1}\sum_{i'}\ket{i'}\bra{i'},\\
\Exp_{P\in \cT} P\ket{i}\bra{j}P^\dagger &=&
  {1\over (N-1)(N-2)}\sum_{i'\not=j'}\ket{i'}\bra{j'},
\een
where the indices in the sums range from $1$ to $N-1$.  This
convention for indices and labels will be in place for the remainder
of the appendix unless otherwise indicated.  The relevant part of
the trivial eigenspace of $\pi_1$ is spanned by $\check I\defeq
\sum_{i'\geq 1}\ket{i'}\bra{i'}$ and an operator with no diagonal
entries.  For $i\not=j$,
\ben
\Exp_{P\in \cT} P\ket{i}\bra{i}P^\dagger\tensor P\ket{i}\bra{i}P^\dagger
   &=& {1\over N-1}\sum_{i'}\ket{i'}\bra{i'}\tensor
	\ket{i'}\bra{i'},\\
\Exp_{P\in \cT} P\ket{i}\bra{i}P^\dagger\tensor P\ket{j}\bra{j}P^\dagger
   &=& {1\over (N-1)(N-2)}\sum_{i'\not=j'}\ket{i'}\bra{i'}\tensor\ket{j'}\bra{j'},\\
\Exp_{P\in \cT} P\ket{i}\bra{j}P^\dagger\tensor P\ket{j}\bra{i}P^\dagger
   &=& {1\over (N-1)(N-2)}\sum_{i'\not=j'}\ket{i'}\bra{j'}\tensor\ket{j'}\bra{i'},\\
\Exp_{P\in \cT} P\ket{0}\bra{j}P^\dagger\tensor P\ket{j}\bra{0}P^\dagger
   &=& {1\over (N-1)}\sum_{j'}\ket{0}\bra{j'}\tensor\ket{j'}\bra{0},\\
\Exp_{P\in \cT} P\ket{j}\bra{0}P^\dagger\tensor P\ket{0}\bra{j}P^\dagger
   &=& {1\over (N-1)}\sum_{j'}\ket{j'}\bra{0}\tensor\ket{0}\bra{j'}
\een
and expressions involving other combinations
of indices which will be of no further concern.
The relevant part of the trivial eigenspace of $\pi_2$
is therefore spanned (non-orthogonally) by
\ben
\check D&\defeq&
   \sum_{i'}\ket{i'}\bra{i'}\tensor\ket{i'}\bra{i'},\\
\check E&\defeq& 
   \sum_{i',j'}\ket{i'}\bra{i'}\tensor\ket{j'}\bra{j'},\\ 
\check J&\defeq&
   \sum_{i',j'}\ket{i'}\bra{j'}\tensor\ket{j'}\bra{i'},\\
\check Z_1&\defeq&
   \sum_{i'}\ket{0}\bra{i'}\tensor\ket{i'}\bra{0},\\
\check Z_2&\defeq&
   \sum_{i'}\ket{i'}\bra{0}\tensor\ket{0}\bra{i'}.
\een

Define
\ben
\check\rho_0 &\defeq&
   \sum_{i\geq 1}\rho_{0i}\ket{i}\bra{0}+\rho_{i0}\ket{0}\bra{i},\nonumber\\
\check\rho_{\bar 0} &\defeq& \check\rho - \check\rho_0.
\een
If $P$ is a random product of operators in $\cT$ and in
a diagonal group $\cD$, then
\bea
\Exp_{P_1\in D,P_2\in \cT} P_2P_1\check\rho P_1^\dagger P_2^\dagger &=& 
   {1\over N-1}\sum_i\check\rho_{ii} \check I = 0,\\
\Exp_{P_1\in D,P_2\in \cT} P_2P_1\check\rho P_1^\dagger P_2^\dagger\tensor
   P_2P_1\check\rho P_1^\dagger P_2^\dagger
   &=&
      {1\over N-1}\sum_{i}\check\rho_{ii}^2\check D\nonumber\\
   &&\;\;+\;\;
      {1\over (N-1)(N-2)}\sum_{i\not= j}
           \check\rho_{ii}\check\rho_{jj}(\check E-\check D)\nonumber\\
   && \;\;+\;\;
      {1\over (N-1)(N-2)}\sum_{i\not=j}
            \check\rho_{ij}\check\rho_{ji}(\check J-\check D)\nonumber\\
   &&\;\;+\;\;
      {1\over N-1}\sum_{i}\check\rho_{0i}\check\rho_{i0}
        (\check Z_1+\check Z_2)\nonumber\\
   &=&
      {1\over N-1}\trace(\check\rho_d^2)\check D\nonumber\\
   &&\;\;+\;\;
      {1\over (N-1)(N-2)}
   (\trace(\check\rho_{\bar 0})^2-\trace(\check\rho_d^2))(\check E-\check D)\nonumber\\
   && \;\;+\;\;
      {1\over (N-1)(N-2)}(\trace(\check\rho_{\bar 0}^2) -\trace(\check\rho_d^2))
   (\check J-\check D)\nonumber\\
   &&\;\;+\;\;
      {1\over 2(N-1)}\trace(\check\rho_0^2)
        (\check Z_1+\check Z_2).
\label{equation:two-trans}
\eea
The variance $\bar v$ is obtained by taking the inner product
of this expression with $\sigma\tensor\sigma$.
Define
\ben
\check\sigma_d &\defeq& \sum_{i}\sigma_{ii}\ket{i}\bra{i},\\
\check\sigma_0 &\defeq& 
\sum_{i}\sigma_{0i}\ket{i}\bra{0}+\sigma_{i0}\ket{0}\bra{i},\\
\check\sigma_{\bar 0} &\defeq& \sigma-\check\sigma_0 - \sigma_{00}\ket{0}\bra{0}.
\een
We will make use of the following (in)equalities:
\ben
\trace\check\rho &=& \trace\check\rho_{\bar 0}\nonumber\\
    &=& \trace\check\rho_d\nonumber\\
    &=&0,\\
\trace(\rho^2) &=& \trace(\check\rho_{\bar 0}^2) + \trace(\check\rho_0^2)+
 \rho_{00}^2 + (N-1)\bar p^2\nonumber\\
               &\leq& 1,\\
\trace(\check\sigma_{\bar 0}) &=& -\sigma_{00},\\
\trace(\sigma^2) &=&
     \trace(\check\sigma_{\bar 0}^2) +\trace(\check\sigma_0^2) + \sigma_{00}^2\nonumber\\
  &=& N,\\
\trace(\check\sigma_0^2) + 2\sigma_{00}^2 &=& 2,\\
\trace(\check\sigma_d^2)&\leq& \trace(\check\sigma_{\bar 0}^2)\nonumber\\
  &\leq&  N-1,
\een
where we used  the
properties of the trace inner product and
the fact that $\sigma$ is unitary.
The variance can now be estimated by
\ben
\bar v &=& 
      {1\over N-1}\trace(\check\rho_d^2)\trace(\check\sigma_d^2)\nonumber\\
   &&\;\;+\;\;
      {1\over (N-1)(N-2)}
   (\trace(\check\rho_{d})^2-\trace(\check\rho_d^2))
   (\trace(\check\sigma_d)^2 - \trace(\check\sigma_d^2))\nonumber\\
   && \;\;+\;\;
      {1\over (N-1)(N-2)}(\trace(\check\rho_{\bar 0}^2) -\trace(\check\rho_d^2))
   (\trace(\check\sigma_{\bar 0}^2) - \trace(\check\sigma_d^2)))\nonumber\\
   &&\;\;+\;\;
    {1\over 2(N-1)}\trace(\check\rho_0^2)\trace(\check\sigma_0^2)\nonumber\\
   &\leq&
     \trace(\check\rho_d^2)
     +{1\over N-2} \trace(\check\rho_d^2)
     +{1\over N-2}(\trace(\check\rho_{\bar 0}^2)-\trace(\check\rho_d^2))
     +{1\over N-1}\trace(\check\rho_0^2)\nonumber\\
   &\leq&
     \trace(\check\rho_d^2) +{1\over N-2}\trace(\check\rho^2).
\een
Both of the terms in this expression can be large compared
to $\bar r^2$. The presence of the second term shows the importance
of ensuring that $\rho$ is initially in a nearly diagonal form,
and implies a limit on the effectiveness of the diagonal
group. However, if $\sigma$ is diagonal
in the computational basis, the second term does not
arise.

The signal to noise ratio for the thermal distribution can
now be obtained as follows. With the definitions from
Section~\ref{section:flip&swap},
\ben
\trace(\check\rho^2) &=&\sum_{i=1}^{N-1}(\rho_{ii}-\bar p)^2\nonumber\\
   &\leq&\sum_{i=0}^{N-1}(\rho_{ii} - {1\over N})^2\nonumber\\
   &=&\sum_{i=0}^{N-1}\rho_{ii}^2 - {1\over N}\nonumber\\
   &=&\prod_{i=1}^n{1\over 4}((1+\delta_i)^2 + (1-\delta_i)^2) - {1\over N}\nonumber\\
   &=&{1\over 2^n}(\prod_{i=1}^n(1+\delta_i^2) - 1)\nonumber\\
   &\leq& {1\over 2^n}(e^{\sum_i\delta_i^2} - 1),\\
\trace\check\rho^2
   &\geq& {1\over 2^n}\sum_i\delta_i^2.
\een
The last expression is a good approximation as long
as $\sum_i\delta_i^2 \ll 1$.
The probability of the ground state is given by
\ben
\rho_{00} &=& {1\over 2^n}\prod_{i=1}(1+\delta_i)\nonumber\\
          &\geq& {1\over 2^n}(1+\sum_i\delta_i),
\een
which is a good approximation as long as $\sum_i\delta_i\ll 1$.
Thus the signal to noise ratio for randomization
using a two-transitive group is bounded by
\bea
	\mbox{SNR} &\geq& {\sum_i\delta_i|\sigma_{00}|\over\sqrt{
		2^{2n} s + 2^n(1+ 1/(2^n-2))\sum_i\delta_i^2}}. 
\eea
To understand the behavior of this expression, consider
the case where $\delta_i = \delta$ is independent of the
qubit. We express $s$ in terms of the signal to noise
ratio $\mbox{SNR}_1$ for a single qubit,
 $\mbox{SNR}_1 \defeq \delta/\sqrt{s}$.
For a typical NMR experiment with protons, $\mbox{SNR}_1 \sim 10^3$.
With these definitions,
\bea
\mbox{SNR} &\geq& {n|\sigma_{00}|\delta\over
                 \sqrt{2^{2n}\delta^2/\mbox{SNR}_1^2 + 2^n(1+
                 1/(2^n-2))n\delta^2}} \nonumber \\ 
&\geq& {n\over 2^n}{\mbox{SNR}_1|\sigma_{00}|\over \sqrt{1 +
                 n\mbox{SNR}_1^2/(2^n-2)}}.
\eea
For small $n$, $\mbox{SNR}$ is dominated by the contribution
to the variance from the randomization process, while for large
$n$, it is dominated by the reduction in excess probability
in the ground state.

\subsection{Cyclic Permutation Groups} 

Consider using a cyclic group $\cS_1$ of permutations which
leave $\ket{0}$ fixed. This was done for exhaustive
averaging, but can also be applied to randomization.
As we will see, the main problem is that the variance
of the measurements cannot be guaranteed to be sufficiently
small. Let $\pi$ be a generator of the group
of order $N-1$.

The trivial eigenspaces of $\cS_1$
can be computed as in the previous section. The relevant subspaces are
spanned by $\check I$ for $\pi_1$ and
\ben
\check D_k &\defeq& \sum_{i\geq
  1}\ket{i}\bra{i}\tensor\ket{\pi^k(i)}\bra{\pi^k(i)},\\
\check J_k&\defeq&
   \sum_{i\geq 1}\ket{i}\bra{\pi^k(i)}\tensor\ket{\pi^k(i)}\bra{i},
\een
$\check Z_1$, $\check Z_2$ and a few others of no further
concern for $\pi_2$.

Let $P$ be a random product of an element of a diagonal group $\cD$
and the cyclic group $\cS_1$.
\ben
\Exp_{P_1\in \cD, P_2\in \cS_1}
   P_2P_1\check\rho P_1^\dagger P_2^\dagger &=& 0,\\
\Exp_{P_1\in \cD, P_2\in \cS_1}
 P_2P_1\check\rho P_1^\dagger P_2^\dagger\tensor P_2P_1\check\rho P_1^\dagger P_2^\dagger 
  &=&
  \sum_{k=0}^{N-2}{1\over N-1}
     \sum_{i}\check\rho_{ii}\check\rho_{\pi^k(i)\pi^k(i)} \check D_k \nonumber\\
  &&+\;\;
  \sum_{k=1}^{N-2}{1\over N-1}
     \sum_{i}\check\rho_{i\pi^k(i)}\check\rho_{\pi^k(i)i} \check J_k \nonumber\\
  &&+\;\;
  {1\over 2(N-1)}\trace(\check\rho_0^2) (\check Z_1+\check Z_2).
\een
To compute $\bar v$ we take the trace after multiplying
by $\sigma\tensor\sigma$. 
\ben
\bar v &=&
  \sum_{k=0}^{N-2}{1\over N-1}
     \sum_{i}\check\rho_{ii}\check\rho_{\pi^k(i)\pi^k(i)}
     \sum_{i}\sigma_{ii}\sigma_{\pi^k(i)\pi^k(i)}\nonumber\\
  &&+\;\;
  \sum_{k=1}^{N-2}{1\over N-1}
     \sum_{i}\check\rho_{i\pi^k(i)}^2
     \sum_{i}\sigma_{i\pi^k(i)}\sigma_{i\pi^k(i)}\nonumber\\
  &&+\;\;
  {1\over N-1}\trace(\check\rho_0^2)\trace(\check\sigma_0^2).
\een
The sum involves off-diagonal expressions and products
of correlations of the diagonals of $\rho$ and $\sigma$.
Although $\bar v$ can be much too high in the worst
case, in practice one can expect it to be close
to what was obtained for a two-transitive group.
However, since the known algorithms for the cyclic groups
are no more efficient than those for the linear group,
there is presently little to be gained by using cyclic
groups.

\subsection{ The Unitary Group}

A spanning set of eigenvectors of the representation $\pi_2$
of the unitary group $U$ acting on $\ket{1},\ldots,\ket{N-1}$
consists of $\check E$, $\check J$, $\check Z_1$, $\check Z_2$
and $\ket{0}\bra{0}\tensor\ket{0}\bra{0}$. As a result
one obtains
\ben
\Exp_{P\in U}P\check\rho P^\dagger\tensor P\check\rho P^\dagger
 &=&
  {1\over 2N(N-1)}
    ((\trace\check\rho_d)^2 + \trace(\check\rho_{\bar 0}^2))(\check E+\check J)\nonumber\\
   &&\;\;+
  {1\over 2(N-1)(N-2)}
     ((\trace\check\rho_d)^2 - \trace(\check\rho_{\bar 0}^2))(\check E-\check J)\nonumber\\
  &&\;\;+{1\over 2(N-1)}\trace(\check\rho_0^2)(\check Z_1 + \check Z_2)\nonumber\\
 &=&
  {1\over N(N-2)}\trace(\check\rho_{\bar 0}^2)\check J
  -{1\over N(N-1)(N-2)}\trace(\check\rho_{\bar 0}^2)\check E
  +{1\over 2(N-1)}\trace(\check\rho_0^2)(\check Z_1 + \check Z_2)
.
\een
Thus
\ben
\bar v &=&
  {1\over N(N-2)}\trace(\check\rho_{\bar 0}^2)\trace(\check\sigma_{\bar 0}^2)
  -{1\over N(N-1)(N-2)}\trace(\check\rho_{\bar 0}^2)\trace(\check\sigma_d)^2
  +{1\over 2(N-1)}\trace(\check\rho_0^2)\trace(\check\sigma_0^2)\nonumber\\
       &\leq&
       {N-1\over N(N-2)}\trace(\check\rho_{\bar 0}^2)
   +{1\over N-1}\trace(\check\rho_0^2)\nonumber\\
       &\leq&
   {1\over N-2}\trace(\check\rho^2).
\een
By using the unitary group, it is possible to eliminate the
term $\trace\check\rho_d^2$ that occurs in the expression
for $\bar v$ for the two-transitive permutation groups.
Although it is impossible to efficiently implement random
elements of the unitary group, there are effective methods
for accomplishing the same by using the normalizer group.

\subsection{The Normalizer Group}

The normalizer group is as effective at randomizing
$\ket{0},\ldots,\ket{N-1}$ as the full unitary group,
at least in terms of expectations and variance.
It is straightforward to determine the trivial eigenspaces
of $\pi_1$ and $\pi_2$ in the Pauli operator basis. For
$i\not=j$ and $j\not=0$,
\ben
\Exp_{P\in \cal N} P\sigma_i P^\dagger &=& \delta_{i,0}\sigma_0,\\
\Exp_{P\in\cal N} P\sigma_i P^\dagger\tensor P\sigma_j P^\dagger
 &=& 0,\\
\Exp_{P\in\cal N} P\sigma_j P^\dagger\tensor P\sigma_j P^\dagger
 &=& {1\over 2^{2m}-1}\sum_{j'\not=0}\sigma_{j'}\tensor\sigma_{j'},\\
\een
where $m$ is the number of qubits.  Using these identities,
it can be verified that the trivial eigenspace of $\pi_1$
is spanned by the identity, and that of $\pi_2$ 
by $E\defeq I\tensor I$ and $J\defeq
\sum_{i,j\geq 0}\ket{i}\bra{j}\tensor\ket{j}\bra{i}$.

To exploit the normalizer group without removing the
polarization in $\ket{0}$ requires conditioning it
on one of the qubits.

\subsection{Conditional Normalizer Group}

In this section we analyze the behavior of the
algorithm based on alternate randomizations using
$\cT$ and the conditional normalizer group $\cN_1$.

Let $\bar R_{k-1}$ be the expectation of $P\check\rho P^\dagger\tensor
P\check\rho P^\dagger$ after the $k$'th step of the conditional
normalizer group algorithm. Using Eq.(\ref{equation:two-trans}),
\ben
\bar R_0 &=&
      {1\over (N-1)(N-2)}\left(N\trace(\check\rho_d^2) 
                         - \trace(\check\rho_{\bar 0}^2)\right)\check D
     +{1\over (N-1)(N-2)}(\trace(\check\rho_{\bar 0}^2)-\trace(\check\rho_d^2))\check J\nonumber\\
     &&\;\;-\;\;{1\over (N-1)(N-2)}\trace(\check\rho_d^2)\check E\nonumber\\
     &&\;\;+\;\;
       {1\over 2(N-1)}\trace(\check\rho_0^2)(\check Z_1+\check Z_2).
\een
Define $\alpha_k$, $\beta_k$, $\gamma_k$ and $\delta$ by $\bar R_k
\defeq \alpha_k \check D + \beta_k\check J + \gamma_k \check E +
\delta (\check Z_1+\check Z_2)$, where we have used the fact that
$(\check Z_1+\check Z_2)$ are not affected by randomization with $\cT$
and $\cN_1$. 

Because $\cN_1$ distinguishes the state of the first qubit, we need to
subdivide the tensors in the expression for $\bar R_0$.  Write
$\check D = \check D_0 + \check D_1$, $\check J = \check J_{00}
+\check J_{01} + \check J_{10} + \check J_{11}$ and $\check E = \check
E_{00} +\check E_{01} + \check E_{10} + \check E_{11}$. 
For
example, $\check D_0 =
\sum_{i=1}^{N/2-1}\ket{i}\bra{i}\tensor\ket{i}\bra{i}$, $\check J_{01} =
\sum_{i=1}^{N/2-1}\sum_{j=N/2}^{N-1}\ket{i}\bra{j}\tensor\ket{j}\bra{i}$
and $\check E_{10} =
\sum_{i=N/2}^{N-1}\sum_{j=1}^{N/2-1}\ket{i}\bra{i}\tensor\ket{j}\bra{j}$,
where we are using the convention that the indices $i\geq 2^{n-1} = N/2$
are those referring to states with the first qubit in state $\ket{1}$.
Randomizing over $\cN_1$ preserves all but one of these expressions:
\ben
\Exp_{P\in\cN_1}\pi_2(P)(\check D_1) &=&
   {2\over N+ 2}\check J_{11} + {2\over N+2}\check E_{11}.\\
\een
Hence
\ben
\Exp_{P\in\cN_1}\pi_2(P)(\bar R_k) &=&
  \alpha_k\check D_0 + \beta_k \check J+{2\over N+2}\alpha_k\check J_{11}
  + \gamma_k\check E +{2\over N+2}\alpha_k\check E_{11} 
  + \delta(\check Z_1+\check Z_2).
\een
Randomizing over $\cT$ gives
\ben
\Exp_{P\in \cT}\pi_2(P)(\check D_{0}) &=& {N-2\over 2(N-1)}\check D,\\
\Exp_{P\in \cT}\pi_2(P)(\check E_{11}) &=& {N\over 2(N-1)}\check D +
   {N\over 4(N-1)}(\check E - \check D),\nonumber\\
  &=&{N\over 4(N-1)}\check D + {N\over 4(N-1)}\check E\\
\Exp_{P\in \cT}\pi_2(P)(\check J_{11}) &=& {N\over 2(N-1)}\check D +
   {N\over 4(N-1)}(\check J - \check D)\nonumber\\
  &=&{N\over 4(N-1)}\check D + {N\over 4(N-1)}\check J,
\een
so that
\ben
\bar R_{k+1} &=& \Exp_{P_2\in \cT}\Exp_{P_1\in\cN_1}\pi_2(P_2P_1)(\bar R_k))\\
 &=&
  \left(1-{N^2\over 2(N-1)(N+2)}\right)\alpha_k\check D 
  + \left( \beta_k + {N\over 2(N+2)(N-1)}\alpha_k\right) \check J 
  + \left( \gamma_k + {N\over 2(N+2)(N-1)}\alpha_k\right)\check E
  + \delta(\check Z_1+\check Z_2)).
\een
The variance $\bar v_{k+1}$ after the $k$'th step is given by
\ben
\bar v_{k+1} &=& \trace (\bar R_{k+1} \sigma\tensor \sigma)\\
  &=&
  \alpha_{k+1}\trace(\check\sigma_d^2)
  +\beta_{k+1}\trace(\sigma_{\bar 0}^2)
  +\gamma_{k+1}(\trace\sigma_d)^2
  +\delta \trace(\sigma_0^2).
\een
We can estimate the coefficients as follows:
\ben
\alpha_0\trace(\check\sigma_d^2)&\leq&{N\over N-2}\trace(\check\rho_d^2),\\
\beta_0\trace(\sigma_{\bar 0}^2)& \leq &{1\over N-2}(\trace(\check\rho_{\bar 0}^2) - \trace(\check\rho_d^2)),\\
\gamma_0(\trace\check\sigma_d)^2 &=&
          -{1\over (N-1)(N-2)}\trace(\check\rho_d^2)(\trace\check\sigma_d)^2\nonumber\\
    &\leq& 0,\\
\delta\trace(\check\sigma_0^2) &\leq&{1\over N-1}\trace\check\rho_0^2,\\
\alpha_{k+1}\trace(\check\sigma_d^2)
  &=&{1\over 2}\left(1+{N-2\over (N-1)(N+2)}\right)\alpha_k\trace(\check\sigma_d)^2\nonumber\\
  &\leq&{1\over 2}e^{1\over N+2}\alpha_k\trace(\check\sigma_d^2)\nonumber\\
  &\leq&{1\over 2^k}e^{k\over N+2}{N\over N-2}\trace(\check\rho_d^2).
\een
Define $\lambda = {e^{1\over N+2}/2}$.
The coefficients $\beta_{k}$ and $\gamma_k$ are monotonically
increasing. The limiting values are
\ben
\beta_{\infty}\trace(\check\sigma_{\bar 0}^2) &=&
          (\beta_0 + {1\over N}\alpha_0)\trace(\check\sigma_{\bar 0}^2)\nonumber\\
   &\leq& {1\over N-2}\trace(\check\rho_{\bar 0}^2),\\
\gamma_{\infty}\trace(\check\sigma_d)^2&=&
          ({1\over  N}\alpha_0+\gamma_0)\trace(\check\sigma_d)^2\nonumber\\
   &\leq& 0.
\een
Thus
\ben
\bar v_{k+1}&\leq&
    \lambda^k{N\over N-2}\trace(\check\rho_d^2)
        +{1\over N-2}\trace(\check\rho^2).
\een
By choosing $k$ large enough, the variance can be reduced to
near that obtainable by randomizing
over the whole unitary group. In fact, if
$k$ is chosen so that $\lambda^k \leq 1/(2(N+2))$,
then the maximum contribution to the variance is
$\bar v \leq {2\over N-1}\trace(\check\rho^2)$. 
Consider the case where $\rho$ is diagonal with
$\rho_{00}$ maximal, $c\defeq \sqrt{s}/(\rho_{00}-\bar p)$ and
the output of the algorithm is deterministic (i.e.
$\sigma_{00}^2 = 1$). Then 
${2\over N-1}\check\rho^2\leq 2\bar p(\rho_{00}-\bar p)$
and
\bea
\mbox{SNR} &\geq& {\rho_{00}-\bar p\over \sqrt{s+2\bar p(\rho_{00}-\bar p)}}
\nonumber
\\
   &\geq& {\sqrt{\rho_{00}-\bar p}\over 
            \sqrt{c^2 \rho_{00} + 2\bar p}}.
\eea
Consequently, if $\bar p\ll c\rho_{00}$, the signal to noise ratio
is dominated by ${1\over c}$, the term due to measurement
noise. If $\bar p \gg c\rho_{00}$, then the signal to noise ratio
is determined by the contribution from the randomization method. As
long as $\bar p$ is sufficiently smaller than $\rho_{00}$ and $c
\leq 1$ the signal to noise ratio is bounded below by a constant,
which ensures that a small number of experiments are required to
determine whether $\sigma_{00} = 1$ or $\sigma_{00} = -1$.  However,
in the case where $\bar p \sim \rho_{00}$, the signal to noise ratio
can be very small, for example if $\rho_{ii} = 0$ or $\rho_{ii} =
\rho_{00}$ for all $i$. The situation where $\bar p \sim \rho_{00}$ is
small arises in the high temperature limit of NMR quantum computation.
In this case the signal to noise ratio can be estimated as
\bea
\mbox{SNR} &\geq& 
  {n\over 2^n}{\mbox{SNR}_1|\sigma_{00}|\over 
  \sqrt{1+2n\mbox{SNR}_1^2/(2^n(2^{n}-1))}}.
\eea

\subsection{Randomized Flip\&Swap}

For fully randomized flip\&swap, each experimental determination of
the output of the computation consists of two experiments. First a
sequence of $k$ random operators implementing the conditional
normalizer method is chosen.  For the present purposes we choose $k$
so that $\lambda^k \leq 1/2(N+2)$.  Next two experiments are
performed. In the first the chosen sequence of operators is applied
before measuring $\sigma$. In the second, the flip\&swap operation is
used before applying the same sequence of random operators and
measuring $\sigma$. The measurements are added to obtain the desired
answer.

This algorithm behaves exactly like a single randomized
experiment with input $\rho_s$ (Eq.(\ref{equation:def_rho_s})) and measurement
variance $s/2$. The variance of the randomization
is therefore given by
\ben
\bar v &\leq& {2\over N-1}\trace(\check\rho_s^2)\\
       &\leq&  {2 n^2\delta^2\over N^2 (N-1)}.
\een
Substituting in the expression for the signal to noise ratio gives
\bea
\mbox{SNR} &\geq&
{n\over 2^n}
   {\mbox{SNR}_1|\sigma_{00}|\over\sqrt{1/2 +
	2n^2\mbox{SNR}_1^2/2^{2n}(2^n-1))}}, 
\eea
where we have taken into account the fact that two experiments
contributed to the signal.

Instead of using the conditional normalizer group, one
can use any set of permutation operators $\{P_i\}_{i=1}^{N-1}$
with $P_i\ket{0} = \ket{0}$ and $P_i\ket{N-1} = \ket{i}$.
For example, a cyclic linear group can be relabeled
to have this property. Because of the symmetries of $\rho_s$,
this is as effective as using a two-transitive group.
Since $\trace(\check\rho_s^2)\leq n^2\delta^2/N^2$,
\ben
\bar v &\leq& {(2^n-1) n^2\delta^2\over 2^{2n}(2^n-2)}
\een
and
\bea
\mbox{SNR} &\geq&
{n\over 2^n}
  {\mbox{SNR}_1|\sigma_{00}|\over \sqrt{1/2 + n^2\mbox{SNR}_1^2/(2^n(2^n-2))}}.
\eea

\section{Implementations of Temporal Averaging Algorithms}

\subsection{Flip\&Swap}

The implementation of labeled flip\&swap for three qubits
and an ancilla is shown in Figure~\ref{figure:lab-f&s}.
The flip\&swap is the first group of gates,
consisting of a not applied to each qubit, followed by
controlled-nots from the first to each of the other qubits,
an $n-1$-controlled-not conditioned on the last $n-1$ qubits
being $\ket{0}$, and finally a reversal of
the first set of controlled-nots. Efficient
quantum networks for the $n-1$-controlled not (generalized
Toffoli gates) are given in~\cite{barenco:qc1995a}.
Note that for diagonal initial states, phase variants are
equivalent, so we can use an $\mbox{SU}$ variant of
the Toffoli gate to avoid ancillas while still having
an $O(n)$ implementation. Also, the computation can
be arranged so that it is $O(n)$ even if controlled
operations can only be performed between adjacent qubits
in a linear ordering.

An efficient method for implementing randomized flip\&swap
 is to choose for
each $\ket{b}\not=\ket{\mbox{\bf 0}}$ an ``easy'' linear operator $L$
modulo 2 such that $L\mbox{\bf 1} = b$.  If $b$ has $w$ one's, such an
operator with at most $n-w$ off-diagonal ones exists. The
corresponding unitary operator in the group of linear permutations can
be implemented with $n-w$ controlled nots.

\subsection{The Normalizer Group}

Every element of the normalizer group $\cN$ operating on $n$ qubits
can be implemented by $O(n^2)$ controlled-nots and $\pi/2$ or $\pi$
rotations of single qubits. For the purposes of randomly choosing one
of the members of $\cN$, the natural representation of $U\in\cN$ is
\ben
U:\sigma_b \rightarrow U\sigma_bU^\dagger &=&
 (-1)^{\langle x,b\rangle}i^{f(b,L)}\sigma_{Lb}.
\een
A uniform random element can be obtained by choosing
$x$ and $L$ uniformly subject to $L^TML = M$
(see Section~\ref{section:normalizer-group}).
The vector $x$ is obtained by setting each of the
$2n$ entries of $x$ independently and uniformly
to $0$ or $1$.
To obtain uniformly distributed valid $L$'s
one can construct $L$ column by column.
Write
\ben
M &=& \left[\begin{array}{cc}0&I\\I&0\end{array}\right],
\een
where the entries are $n$ by $n$ matrices and the partitioning is
based on writing the index $b$ of $\sigma_b$ in the form $b = b_0b_1$,
with $b_0$ and $b_1$ containing the indices coming from the first and
second members of each qubit's pair, respectively.

If $L_{\leq k}$ is the $2n$ by $k$ matrix consisting of the
first $k$ columns of $L$, then
$L_{\leq k}^T M L_{\leq k} = M_{\leq k,\leq k}$,
where $M_{\leq k,\leq k}$ is the $k$ by $k$ matrix
submatrix of $M$ in the upper left corner.
The columns of $L_{\leq k}$ are linearly independent
(modulo $2$).
Suppose $L_{\leq k}$ has been constructed and we
wish to add another column to obtain $L_{\leq k+1}$.
The new column $L_{k+1}$ has to satisfy
\ben
L_{k+1}^T M L_{k+1} &=& 0,\\
L_{k+1}^T M L_{\leq k} &=& M_{k+1,\leq k}.
\een
The first equality is satisfied for any $L_{k+1}$, so we wish to
choose $L_{k+1}$ randomly, not in the span of $L_{\leq k}$ and subject
to the second equality. The dimension of the affine space of solutions
to this equality is $2n-k$, while the dimension of the span of
$L_{\leq k}$ is $k$. We consider two cases. If $k< n$, then
$M_{k+1,\leq k} = 0$, and the span of $L_{\leq k}$ is contained in the
space of solutions. Because $2n-k>k$, suitable $L_{k+1}$ can be
found. To pick $L_{k+1}$ uniformly one can use
any algorithm (e.g. one based on Gaussian elimination modulo $2$)
to obtain $2n-2k$ vectors $S_1,\ldots,S_{2n-2k}$ independent of the columns of
$L_{\leq k}$ which together with $L_{\leq k}$ span the solution
space. A random $L_{k+1}$ is obtained by choosing a random
non-zero linear combination of the $S_1,\ldots,S_{2n-2k}$
and adding it to a random linear combination of the columns
of $L_{\leq k}$.

If $k\geq n$, then $M_{k+1,\leq k}$ is non-zero. If $y$ is in the span
of $L_{\leq k}$, then the $k-n+1$'th entry of $y^T M L_{\leq k}$ is
zero. Since that entry of $M_{k+1,\leq k}$ is $1$, the set of
solutions to $y^T M L_{\leq k} = M_{k+1,\leq k}$ does not contain any
element $y$ in the span of $L_{\leq k}$.  We can therefore pick a random
element in this affine subspace of dimension $2n-k$. An affine basis
for this subspace can again be obtained by a Gaussian
elimination method.

The above construction shows that the number of valid
$L$'s is $\prod_{k=0}^{n-1} (2^{2n-k}- 2^k)\prod_{k=0}^{n-1} 2^{n-k}$.
In view of the technique for constructing random invertible
matrices over ${\bf Z}_2$ given in~\cite{randall:qc1993a},
there are probably more efficient methods for constructing
random $L$'s.

To obtain a quantum network which implements the unitary operator
defined by $(x,L)$ requires decomposing $L$ into elementary operations
corresponding to controlled-nots and single qubit rotations.  This can
be done by adapting the methods described in~\cite{cleve:qc1996b}. The
basic idea is to multiply $L$ on the left and right be the linear
operators corresponding to controlled-nots and rotations.  Since
controlled-nots correspond to elementary row/column operations in the
$n$ by $n$ subblocks, one can apply Gaussian elimination methods to
convert the first (say) subblock to standard form.  The $\pi/2$
rotations around the different axes permit elementary row/column
operations between corresponding rows/columns of different
subblocks. This can be used to transform $L$ to $I$. The
representation of the resulting sequence of controlled-nots and
rotations is of the form $(x', L)$. To correct the first
component, one can apply $\sigma_{M(x-x')}$ to the qubits.
The total number of gates needed to implement an
operator in $\cN$ is $O(n^2)$~\cite{cleve:qc1996b}.

\noindent{\bf Implementing $\cD$.}
Being a subgroup of $\cN$, it is clear that each
operator in $\cD$ has an efficient quantum network.
The random phase shifts  of $\cD$ are described by
operators $D(x,B)$ defined by
$D(x,B)(\ket{k})=
i^{\langle x, k\rangle} (-1)^{\langle k, Bk\rangle}$.
A random such operator is obtained by choosing
$x$ randomly and uniformly from all $n$ dimensional
vectors over $\{0,1,2,3\}$ and $B$ uniformly from
the set of strictly upper triangular $n$ by $n$ $0-1$ matrices.
Given such an $x$ and $B$, the phase shifts are implemented
by first applying phase shifts by $i^{x_j}$ of
$\ket{1}$ to the $j$'th qubit, and then
performing a sequence of controlled-sign flips.
The sequence of controlled-sign flips can be read
off the entries of $B$ by the following procedure:
If $B_{ij} = 1$, apply a controlled sign-flip between
bits $i$ and $j$. The number of operations required
to apply the random phase shift is at most
$n(n-1)/2$.

\noindent{\bf Implementing $\cT$.}
A unitary operator $U$ in $\cT$ is defined by $U\ket{b} = \ket{Lb}$
for an invertible (modulo $2$) $n$ by $n$ matrix $L$.  Any such
unitary operator can be implemented using only controlled-nots. Since
a controlled not corresponds to an elementary row/column operation, a
decomposition of $L$ into such operations yields the desired quantum
network. The decomposition can be accomplished by the usual Gaussian
elimination methods. A random invertible $L$ can be generated column
by column using a simpler version of the method described for the
normalizer group.  A more efficient algorithm which can be used to
construct the decomposition into elementary operations at the same
time is described in~\cite{randall:qc1993a}.

\subsection{Entanglement}

The operations $P_1$ and $P_2$ required to implement the
method for effective pure states by entanglement
are implemented as follows. A phase variant
equivalent to $P_1$ for diagonal initial states is obtained
by applying a $\pi/2$ rotation around the $y$ axis to
each of the second group of $n$ qubits. The 
operation $P_2$ is decomposed into the product
of $P_{2,i}\ket{a}\ket{b}\rightarrow\ket{a x^{b_i 2^i}}\ket{b}$
for $i=0,\ldots,n-1$. Multiplication by $x^{b_i 2^i}$
in $\mbox{GF}(2^n)$ is a linear map modulo two and
defines an element of $\cT$ which can be implemented
with $O(n^2)$ controlled-nots. Each $P_{2,i}$ can therefore
be implemented with $O(n^2)$ Toffoli gates, and $P_2P_1$
takes $O(n^3)$ operations.


\end{document}